\documentclass[twocolumn]{aastex631}

\shorttitle{Reconciling Reionization with JWST}
\shortauthors{Bera et al.}
\date{Accepted: 2026 April 11; Revised: 2026 April 9; Received: 2025 December 6}

\usepackage{xspace}
\usepackage{wasysym}    % Extra astronomy symbols
\usepackage{xcolor}
\usepackage{hyperref}

\newcommand{\eor}{${\bf EoR}$}
\newcommand{\eorphisf}{{\bf EoR-$\phi_{\rm UV}$-sf }}
\newcommand{\eorrhosf}{{\bf EoR-$\rho_{\rm UV}$-sf }}
\newcommand{\eorphiwf}{{\bf EoR-$\phi_{\rm UV}$-wf }}
\newcommand{\eorrhowf}{{\bf EoR-$\rho_{\rm UV}$-wf }}
\begin{document}

\title{Towards Reconciling Reionization with JWST: The Role of Bright Galaxies and Strong Feedback}

\author[0000-0001-7072-570X]{Ankita Bera}
\altaffiliation{CTC Post-Doctoral Fellow}
\affiliation{Department of Astronomy, University of Maryland, College Park, MD 20742}
%\correspondingauthor{Ankita Bera}
\email{E-mail: ankitabm@umd.edu}

\author[0000-0002-1050-7572]{Sultan Hassan}
\affiliation{Space Telescope Science Institute, 3700 San Martin Dr, Baltimore, MD 21218}

\author[0000-0002-1109-1919]{Robert Feldmann}
\affiliation{Department of Astrophysics, University of Zurich, Zurich CH-8057, Switzerland}

\author[0000-0002-1109-1919]{Romeel Davé}
\affiliation{Institute for Astronomy, University of Edinburgh, Royal Observatory, Blackford Hill, Edinburgh EH9 3HJ, UK}
\affiliation{Department of Physics and Astronomy, University of the Western Cape, Bellville, 7535 Cape Town, South Africa}

\author[0000-0002-0496-1656]{Kristian Finlator}
\affiliation{New Mexico State University, MSC 4500, PO BOX 30001, Las Cruces, NM 88003}
\affiliation{Cosmic Dawn Center (DAWN), Niels Bohr Institute, University of Copenhagen / DTU-Space
Technical University of Denmark}

%%%%%%%%%%%%%%%%%%%%%%%%%%%%%%%%%%%%%%%%%%%%%%%%%%%%%%%%%%%%%%%%%%%%%%%%
\begin{abstract}

The elevated UV luminosity functions (UVLF) from recent James Webb Space Telescope (JWST) have challenged the viability of existing theoretical models. To address this, we use a semi-analytical framework -- which couples a physically motivated source model derived from radiative-transfer hydrodynamic simulations of reionization with a Markov Chain Monte Carlo sampler -- to perform a joint calibration to JWST galaxy surveys (UVLF, $\phi_{\rm UV}$ and UV luminosity density, $\rho_{\rm UV}$) and reionization-era observables (ionizing emissivity, $\dot{N}_{\rm ion}$, neutral hydrogen fraction, $x_{\rm HI}$, and Thomson optical depth, $\tau$). We find that models with weak feedback and a higher contribution from faint galaxies reproduce the reionization observables but struggle to match the elevated JWST UVLF at $z > 9$. In contrast, models with stronger feedback (i.e., rapid redshift evolution) and a higher contribution from bright galaxies successfully reproduce JWST UVLF at $z \geq 10$, but over-estimate the bright end at  $z < 9$. The strong-feedback model constrained by JWST UVLF predicts a more gradual and extended reionization history, as opposed to the sudden reionization seen in the weak-feedback models. This extended nature of reionization ($z\sim 16$ - $6$) yields an optical depth consistent (at 2-$\sigma$) with the Cosmic Microwave Background (CMB) constraint, thereby alleviating the photon-budget crisis. In both scenarios, reionization is complete by $z \sim 6$, consistent with current data. Our analysis highlights the importance of accurately modeling feedback and ionizing emissivities from different source populations during the first billion years after the Big Bang.

\end{abstract}
%%%%%%%%%%%%%%%%%%%%%%%%%%%%%%%%%%%%%%%%%%%%%%%%%%%%%%%%%%%%%%%%%%%%%%%%

%%%%%%%%%%%%%%%%%%%%%%%%%%%%%%%%%%%%%%%%%%%%%%%%%%%%%%%%%%%%
\section{Introduction} \label{sec:intro}
%%%%%%%%%%%%%%%%%%%%%%%%%%%%%%%%%%%%%%%%%%%%%%%%%%%%%%%%%%%%
%
The recent observations with the James Webb Space Telescope (JWST) have revolutionized our understanding of galaxy formation in the early Universe, pushing UV luminosity function ($\phi_{\rm UV}$) measurements to redshifts as high as $z \sim 14$ using surveys, such as, the GLASS-JWST Early Release Science \citep{2022ApJ...935..110T}, the JWST Advanced Deep Extragalactic Survey (JADES; \citealt{2023arXiv230602465E}), the Cosmic Evolution
Early Release Science field (CEERS; \citealt{2024ApJ...969L...2F}), the COSMOS-Web \citep{2024ApJ...965...98C}, the Prime Extragalactic Areas for Reionization and Lensing Science (PEARLS; \citealt{2023AJ....165...13W}), the Next Generation Deep Extragalactic Exploratory Public survey (NGDEEP; \citealt{2024ApJ...965L...6B}). These observations reveal a steep faint-end slope ($\alpha \lesssim -2.3$) and a striking overabundance of bright galaxies compared to pre-JWST $\Lambda$CDM predictions \citep[e.g.,][]{2023MNRAS.523.1009B, 2023ApJS..265....5H, 2024ApJ...969L...2F, 2024ApJ...965...98C, 2024MNRAS.527.5004M, 2024ApJ...966...74W, 2023MNRAS.518.6011D, 2024ApJ...970...31R}. Integrating the UVLF down to a limiting magnitude (often rest-frame $M_{\rm UV} \sim -17$) yields the UV luminosity density $\rho_{\rm UV}$, a key tracer of the cosmic star formation rate density and ionizing photon production rate (\citealt{2015ApJ...802L..19R, 2015ApJ...803...34B}).

However, interpreting these measurements in the context of reionization models introduces significant tensions. The epoch of reionization marks the last major phase transition of the Universe, when the intergalactic medium changed from cold and neutral at $z \sim 30$ \citep[e.g.,][]{2001PhR...349..125B, 2019arXiv190706653W} to hot and ionized by $z \sim 5$ \citep[e.g.,][]{2020MNRAS.491.1736K, 2022MNRAS.514...55B}. Although this transformation is well established, the timing and topology of reionization yet to be uncovered \citep[for reviews see e.g.][]{2001PhR...349..125B, 2019arXiv190706653W, 2021arXiv211013160R}. Moreover, the unexpectedly elevated UVLF and $\rho_{\rm UV}$ at high redshift has spurred a large number of theoretical and observational efforts and many possible solutions have been proposed. For instance, a rapid buildup of stellar mass at canonical star formation efficiencies, which is difficult to reproduce in current galaxy formation simulations \citetext{e.g., \citealp{2023ApJ...946L..13F}; though see \citealp{2025arXiv250821126S}}, and challenges $\Lambda$CDM-based models of halo assembly; a higher star formation efficiency in early halos potentially reaching $\gtrsim 10\%$ can produce more luminous galaxies at fixed halo mass (\citealt{2023MNRAS.519..843M}). This is supported by semi-analytical models tuned to JWST data that require elevated efficiency in halos of mass $M_{\rm h} \sim 10^{10-11}\, M_\odot$ to match the observed UVLFs. Additionally, bursty star formation histories can lead to UV variability, causing more galaxies to appear in a high-luminosity phase at the time of observation. \citet{2023MNRAS.526.2665S} and \citet{2023MNRAS.525.3254S} show that including stochastic star formation (with scatter of $\sim 0.3\,  {\rm dex}$) helps recover the observed bright-end excess without invoking exotic physics. {Additionally, a top-heavy or evolving stellar initial mass function (IMF), 
characterized by a flatter high-mass slope that shifts the effective mass 
distribution toward more massive stars, has been suggested as a mechanism 
to boost the ionizing photon production efficiency per unit stellar mass 
formed, leading to brighter UVLFs for a given halo mass \citep{2022MNRAS.514.4639C, 2024MNRAS.530.2453C, 2025A&A...694A.254H}.} Other possibilities include reduced dust attenuation, especially in low-metallicity early galaxies (\citealt{2024A&A...684A.207F, 2023MNRAS.526.2196M}), or enhanced merger-driven starbursts, though the latter remains speculative. 

These elevated UVLFs significantly impact inferred reionization histories. For instance, integrating UVLFs with assumed ionizing photon production efficiencies ($\xi_{\rm ion}$) and escape fractions ($f_{\rm esc}$) yields ionizing emissivity values that may overshoot CMB-derived electron scattering optical depth constraints (e.g., Planck 2018; \citealt{2020A&A...641A...6P}) \citep{2024MNRAS.535L..37M}. Moreover, models that fit the UVLF alone often fail to match observed galaxy clustering (i.e., galaxy bias) at high redshift. \citet{2025arXiv250307590C} showed that resolving this tension requires introducing a mass-dependent duty cycle and allowing star formation timescales to evolve with halo mass and redshift.

While JWST has dramatically improved our census of high-redshift galaxies, interpreting UVLF and $\rho_{\rm UV}$ in isolation can bias our understanding of the reionization timeline. Inferring the timeline of cosmic reionization remains one of the most challenging tasks due to the diversity of available probes, each sensitive to different aspects of the ionization state of the intergalactic medium (IGM). Measurements such as the UV luminosity function and UV luminosity density from high-redshift galaxy surveys, especially with JWST, are powerful tools for constraining the ionizing photon budget. However, these probes often rely on uncertain assumptions about quantities like the escape fraction of ionizing photons and the ionizing efficiency of stellar populations. Consequently, models tuned solely to match $\phi_{\rm UV}$ or $\rho_{\rm UV}$ can over-predict or under-predict the ionizing emissivity or suggest unrealistically early or rapid reionization histories. On the other hand, low-redshift, global constraints, such as the volume-averaged neutral hydrogen fraction, $x_{\rm HI}(z)$ from quasar damping wings \citep[e.g.,][]{2018ApJ...864..142D, 2019MNRAS.484.5094G, 2020ApJ...904...26Y} or Ly$\alpha$ emitter statistics \citep[e.g.,][]{2008ApJ...677...12O, 2012ApJ...744...83O, Mason_2018, 2020ApJ...904..144J, 2024ApJ...967...28N}, and the CMB optical depth, $\tau$ \citep{2020A&A...641A...6P} are more directly sensitive to the integrated effect of ionizing sources over time, but lack information about the underlying galaxy populations responsible. Each of these observables, when considered in isolation, carries inherent degeneracies and can yield significantly different reionization histories, particularly in the timing, duration, and steepness of the ionization transition. 

In this paper, we explore broad range of reionization scenarios using a flexible framework that jointly incorporates high-redshift galaxy and low-redshift IGM observables. By combining these complementary datasets across redshift and methodology we try to break degeneracies, reconcile apparent tensions, and develop a consistent picture of the sources, timing, and topology of cosmic reionization.
This paper is organized as follows: we describe our source model in \S\ref{sec:source}. In \S\ref{sec:obs}, we outline the observational datasets, specifically, the JWST UV luminosity function and UV luminosity density datasets are presented in \S\ref{subsubsec:data_uvlf} and \S\ref{subsubsec:data_rhouv}, respectively, along with the EoR datasets described in \S\ref{subsubsec:data_nion}, \S\ref{subsubsec:data_xHI}, and \S\ref{subsubsec:data_tau}. The calibration of different models are described in \S\ref{sec:model_calib}. We then jointly constrain our models using the JWST $\phi_{\rm UV}$ and $\rho_{\rm UV}$ together with reionization observables in \S\ref{res:uvlf_res}, \S\ref{res:rho_res}, and \S\ref{res:eor_res}. Finally, we present our concluding remarks in \S\ref{sec:conc}.

We assume a flat, $\Lambda$CDM cosmology throughout this paper with the cosmological parameters obtained from the recent \citet{2020A&A...641A...6P} measurements, i.e. $\Omega_{\Lambda} = 0.69$, $\Omega_{\rm m} = 0.31$, $\Omega_{\rm b} = 0.049$, and the Hubble parameter $H_0 = 67.66$~km/s/Mpc.

%%%%%%%%%%%%%%%%%%%%%%%%%%%%%%%%%%%%%%%%%%%%%%%%%%%%%%%%%%%%
\section{Source model} \label{sec:source}
%%%%%%%%%%%%%%%%%%%%%%%%%%%%%%%%%%%%%%%%%%%%%%%%%%%%%%%%%%%
\subsection{Ionization rate} \label{subsec:Rion}

In this section, we provide a brief description of the semi-analytical model utilized in this study. To investigate the impact of different galaxy populations, we adopt a {\it flexible} physically motivated source model originally developed in \citet{2016MNRAS.457.1550H} based on the cosmological hydrodynamic simulations \citep{2013MNRAS.434.2645D} and radiative transfer simulations of reionization presented in \citet{2015MNRAS.447.2526F}, which later has been calibrated to cosmic dawn and reionization constraints in \citet{2023ApJ...959....2B}. This model parameterizes the ionization rate, {$R_{\rm ion} (\rm s^{-1})$}, as a function of both halo mass ($M_h$) and redshift ($z$), and does not include the escape fraction. The latter is introduced when calculating the ionizing emissivity. The mass dependence of $R_{\rm ion}$ follows a Schechter-like function, which exhibits a power-law behavior at the bright end, {reflecting the correlation between halo mass and star formation rate \citep{2011ApJ...743..169F} and an exponential suppression at low masses due to feedback processes such as supernova-driven outflows and photoionization heating. Meanwhile, the redshift dependence is described by a simple power-law relation, that captures the smooth evolution of star formation efficiency and feedback strength with cosmic time, and was shown to provide a good fit to the redshift-dependent emissivity extracted from these simulations \citep{2016MNRAS.457.1550H, 2017MNRAS.468..122H}.} This parameterization effectively accounts for the non-linear dependence of $R_{\rm ion}$ on halo mass, which can be expressed as,
\begin{equation} \label{eq:Rion}
    \frac{R_{\rm ion}[\rm s^{-1}]}{M_h[\rm M_\odot]} = A\,(1+z)^D \left(\frac{M_h}{B}\right)^C \exp\left[-\left(\frac{M_h}{B}\right)^{-3}\right] \, ,
\end{equation}

where $A$, $B$, $C$, and $D$ are free parameters. The parameter $A$ serves as the amplitude of $R_{\rm ion}$, uniformly scaling the ionizing emissivity across all halo masses at a given redshift. The parameter $B$ sets the minimum halo mass that can contribute to ionizing radiation effectively representing a quenching mass scale influenced by feedback from star formation and photoionization heating. The parameter $C$ defines the slope of the $R_{\rm ion}$–$M_h$ relation, shaping how different halo mass ranges contribute to the overall emissivity. Finally, $D$ captures the redshift evolution of the ionization rate for halos of a fixed mass and can also be interpreted as a proxy for the strength of redshift-dependent feedback mechanisms.

With our source model $R_{\rm ion}$ in place, we use the Sheth-Tormen halo mass function (HMF) \citep{1999MNRAS.308..119S}, $\frac{\text{d}n}{\text{d}M}$, which describes the number density of halos per unit mass, to calculate the redshift evolution of key global quantities driving reionization and cosmic dawn. To compute the HMF, we utilise the mass function module from {\it Colossus} \citep{2018ApJS..239...35D}. We consider halos in the mass range $10^{5}$ to $10^{15}\, M_{\odot}$, capturing contributions from all source populations, including the faintest halos. This comprehensive approach ensures that we account for the entire population of halos that may contribute to the ionizing photon budget during the early stages of cosmic evolution.

We incorporate the metallicity dependence of the emissivity using the sub-solar metallicity-dependent parameterization provided by \citet{2011ApJ...743..169F},
\begin{equation} \label{eq:Qion}
    \log Q_{\rm ion}(Z) = 0.639 (-\log Z)^{1/3} + 52.62 \, ,
\end{equation}

where, Q is in units of $s^{-1}\,(M_\odot\,\text{yr}^{-1})^{-1}$, and $Z$ is the metal mass fraction. This fitting function was calibrated to match the equilibrium emissivities tabulated in \citet{2003A&A...397..527S} over the metallicity range $Z \in [10^{-7}, 0.04]$. The two terms on the right hand side capture the enhanced ionizing efficiency of metal-poor stellar populations, which emit more ionizing photons due to their higher effective temperatures. 
%The last term adjusts the normalization to reflect the \cite{2003PASP..115..763C} initial mass function.
To account for the chemical evolution of star-forming galaxies, we adopt a redshift-dependent mass-weighted metallicity ($Z$) relation, as derived in \citet{2017ApJ...840...39M},
\begin{equation}
    \log\left(Z/Z_\odot\right) = 0.153 - 0.074\,z^{1.34} \, .
\end{equation}
%
%-------------------------------------------------%
\subsection{Star formation rate \& UV luminosity density} \label{subsec:sfr_luv}

We infer the star formation rates (SFR) from the ionization rate $R_{\rm ion}$, given in eq.~\ref{eq:Rion} by combining it with a stellar metallicity-dependent parameterization of the ionizing photon production rate, $Q_{\rm ion}(Z)$, given in eq.~\ref{eq:Qion} by,
\begin{equation} \label{eq:SFR}
    {\rm SFR} = R_{\rm ion}(M_h,z)/ Q_{\rm ion}(Z) \, .
\end{equation}

We assume that the SFR is related to the rest-frame UV luminosity of galaxies, $L_{\text {UV}}$ as,
\begin{equation}
    {\rm SFR} [{\rm M_\odot\,yr^{-1}}] = \kappa_{\text{UV}} L_{\rm UV} [{\rm ergs\,s^{-1}\,Hz^{-1}}] \, .
\label{eq:UVtoSFR}
\end{equation}
$\kappa_{\text{UV}}$ is conversion factor, and we adopt the value $\kappa_{\text{UV}} = 1.15 \times 10^{-28}$ from \citet{2014ARA&A..52..415M} as it is consistent with the cosmic star formation history and the evolved metallicity up to $z\sim 10$. However, we note that, the value of $\kappa_{\rm UV}$ depends on the stellar initial mass function (IMF), metallicity, and star formation history (SFH). For a Salpeter IMF and constant SFR, stellar population synthesis models (FSPS and GALAXEV) yield values of $\kappa_{\rm UV}$ ranging from $1.0$ to $1.55 \times 10^{-28} \,\text{M}_\odot\,\text{yr}^{-1}/(\text{erg s}^{-1}\text{Hz}^{-1})$, with lower metallicities producing slightly higher conversion factors. The adopted value here is lower than the classical \citet{1998ARA&A..36..189K} value of $1.4 \times 10^{-28}$. 
{Note that, at higher redshifts, stellar populations tend to have lower metallicities and younger ages, both of which produce hotter, more UV-luminous stars, thereby lowering $\kappa_{\rm UV}$. However, these effects are partially offset by the shorter time available for stellar populations to reach equilibrium and by the evolving shape of the star formation history. The net result is that $\kappa_{\rm UV}$ varies by less than $\sim$20\% over the relevant redshift range ($z \sim 4$--$14$), as shown by stellar population synthesis models \citep{2014ARA&A..52..415M}.}
For alternate IMF choices (e.g., Chabrier or Kroupa), constant rescaling factors (e.g., 0.63 or 0.67) can convert SFRs and stellar masses to Salpeter equivalents.

We obtain the globally averaged UV luminosity density ($\rho_{\rm UV}$) by integrating the UV luminosity over the entire halo mass range at different redshifts as,
\begin{equation}
    \rho_{\rm UV} \, [\rm{erg\,s^{-1}\,Hz^{-1}\,Mpc^{-3}}] = \int L_{\rm UV} \, \frac{{\rm d}n}{{\rm d}M_h} \, {\rm d}M_h \, .
\end{equation}
%-------------------------------------------------%
\subsection{Galaxy UV luminosity functions} \label{subsec:uvlf}
The ultraviolet luminosity function (UVLF) of galaxies quantifies the number density of galaxies as a function of their intrinsic UV luminosity and provides a key observational probe of galaxy formation and evolution, particularly at high redshifts. Since the rest-frame far-UV light in galaxies is dominated by young, massive stars, the UVLF directly traces the star formation activity across cosmic time. In order to calculate the intrinsic UVLF, $\phi_{\rm UV}(M_{\rm UV}, z)$ from UV luminosity, $L_{\rm UV}$, we first relate it to the absolute magnitude, $M_{\rm UV}$ (in the AB magnitude system) using the relation \citep{1983ApJ...266..713O},
\begin{equation} \label{eq:uvlf}
    M_{\rm UV} = -2.5 \log_{10} \left(\frac{L_{\rm UV}}{\rm{erg\,s^{-1}\,Hz^{-1}}} \right) + 51.63 .
\end{equation}

We note that, recent JWST observations confirm a population of intrinsically UV-luminous galaxies ($M_{\rm UV} < -20$; e.g., \citealt{2022ApJ...940L..14N, 2024ApJ...969L...2F, 2023ApJS..265....5H, 2024ApJ...970...31R}) that exhibit very blue UV continuum slopes ($\beta_{\rm UV}$), with indications of dust-free or negligible dust attenuation at $z > 10$ \citep{2024MNRAS.531..997C}. Additionally, the galaxy population by BoRG-JWST survey at $z \sim 8$ appear to be consistent with being dust-poor \citetext{\citealp{2025arXiv250701014R}; also see, \citealp{2024MNRAS.531..997C, 2025MNRAS.540.2081F}}. Although we do not explicitly model dust in our current framework, the presence of dust at early times cannot be completely ruled out \citep{2020MNRAS.492.5167V, 2023Natur.621..267W, 2023MNRAS.526.4801T, 2023ApJ...943L..27F, 2025MNRAS.540.3693S}, and we intend to explore its impact more carefully in future work.

We then bin the absolute magnitude and estimate the number density of galaxies per magnitude bin using the Sheth-Tormen HMF \citet{1999MNRAS.308..119S}, and verify with the analytical prescription,
\begin{equation}
    \phi_{\rm UV} [\rm Mpc^{-3} \, mag^{-1}] = \frac{{\rm d}n}{{\rm d}M_h} \, \left|\frac{{\rm d}M_h}{{\rm d}L_{\rm UV}} \right| \, \left| \frac{{\rm d}L_{\rm UV}}{{\rm d}M_{\rm UV}} \right|
\end{equation}

%-------------------------------------------------%
\subsection{Ionizing emissivity}\label{subsubsec:model_nion}
The integrated emission rate density of ionizing photons, referred to as the ionizing emissivity ($\dot{N}_{\rm ion}$), quantifies the total number of ionizing photons escaping into IGM per second per unit volume. This quantity is directly linked to $R_{\rm ion}$ through the relation,
\begin{equation} \label{eq:Nion}
\dot{N}_{\rm ion} \, [{\rm s^{-1} \, Mpc^{-3}}] = f_{\rm esc} \int R_{\rm ion}(M_h, z) \, \frac{{\rm d}n}{{\rm d}M_h} \, {\rm d}M_h \, ,
\end{equation}
where $\frac{{\rm d}n}{{\rm d}M_h}$ denotes the differential halo mass function from \citet{1999MNRAS.308..119S}, which provides the number density of halos per unit mass in the interval $M_h$ to $M_h + {\rm d}M$ per unit comoving volume. 
While significant progress has been made, the escape fraction ($f_{\rm esc}$) remains one of the most uncertain parameters in reionization modeling, with estimates ranging from a few percent to over $50\%$, depending on galaxy mass, redshift, and evolutionary state \citep{Wise2009, gnedin2008escape, 2011MNRAS.412..411Y, 2010MNRAS.401.2561W, Paardekooper2015, Ma2015, Chisholm2018, Yeh2022, 2023MNRAS.521.3077K, 2017MNRAS.470..224T, 2023MNRAS.523L..35M, 2024MNRAS.529.3751C}. In this work, we assume an effective $f_{\rm esc}$, averaging over the entire population. However, one can interpret the parameters $C$ and $D$ as effectively capturing the mass and redshift dependence of $f_{\rm esc}$, or more precisely, these dependencies convolved with those of $R_{\rm ion}$. We defer a detailed disentanglement of the mass and redshift dependencies of $R_{\rm ion}$ and $f_{\rm esc}$ to future work.
%
%-------------------------------------------------%
\subsection{Neutral hydrogen fraction} \label{subsec:model_xHI}
We compute the reionization history in our models using the following approach. The time evolution of the ionized hydrogen fraction in the intergalactic medium, $x_{\rm HII}$, is given by \citet{1999ApJ...514..648M},
\begin{equation}
\frac{{\rm d}x_{\rm HII}}{{\rm d}t} = \frac{\dot{N}_{\rm ion}}{\bar{n}_{\rm H}} - \frac{x_{\rm HII}}{\bar{t}_{\rm rec}} \, .
\end{equation}

The first term on the right-hand side represents the production rate of ionized hydrogen, expressed as the ratio of the comoving ionizing emissivity $\dot{N}_{\rm ion}$ to the average comoving number density of hydrogen atoms $\bar{n}_{\rm H}$, which is defined as,
\begin{equation}
\bar{n}_{\rm H} = \frac{X \, \Omega_{b,0} \, \rho_{{\rm crit},0}}{m_{\rm H}} ,
\end{equation}

Here, X = 0.76 is the hydrogen mass fraction, $\Omega_{b,0}$ is the present-day baryon density parameter, $\rho_{{\rm crit},0}$ is the critical density today, and $m_{\rm H}$ is the mass of a hydrogen atom.

The second term accounts for the loss of ionizing photons due to recombinations, characterized by the recombination timescale $t_{\rm rec}$, given by,
\begin{equation}
t_{\rm rec} = \left[ C_{\rm HII} \, \alpha_{\rm A} \, (1 + \chi) \, \bar{n}_{\rm H} \, (1+z)^3 \right]^{-1}  .
\end{equation}

In this expression, $C_{\rm HII}$ is the clumping factor of ionized hydrogen, which captures the effect of density inhomogeneities in the IGM and enhances the effective recombination rate. {We adopt the redshift-dependent form of $C_{\rm HII}$ from \citet{2015MNRAS.451.1586P}, {which we approximate by the cubic fit $C_{\rm HII}(z) = 11.822 - 1.579\,z + 0.0804\,z^2 - 0.00148\,z^3$. This predicts a boost of up to a factor of $\sim 5$ near the end of reionization, reflecting the growth of density inhomogeneities in the photoionized IGM as structure formation proceeds.} The helium correction term is $\chi = Y / 4X$, where Y = 0.24 is the helium mass fraction, and X is the cosmic hydrogen mass fraction (0.76). The recombination coefficient $\alpha_{\rm A}$ corresponds to case A recombination at a temperature of $10^4\,{\rm K}$, with a value of $4.2 \times 10^{-13}\,{\rm cm^3\,s^{-1}}$ \citep{Kaurov_2014}.

%-------------------------------------------------%
\subsection{Optical depth} \label{subsec:model_tau}
The Thomson scattering optical depth, $\tau$, quantifies the cumulative scattering of cosmic microwave background (CMB) photons by free electrons generated during reionization. Given a specific reionization history, $\tau$ can be computed as,
\begin{equation}
    \tau = \int_0^{\infty} \rm{d}z\, \frac{c(1+z)^2}{H(z)}\, \sigma_T\, \bar{n}_{\rm H}\, \left[ x_{\rm HII}(1+\chi) + \chi\,x_{\rm HeIII} \right] ,
\end{equation}
where $\sigma_T$ is the Thomson cross-section, c is the speed of light, and $\bar{n}_{\rm H}$ is the comoving hydrogen number density. The terms $x_{\rm HII}$ and $x_{\rm HeIII}$ denote the ionized fractions of hydrogen and doubly ionized helium, respectively, and $\chi = Y / 4X$ accounts for the helium contribution, with X and Y being the hydrogen and helium mass fractions. This integral follows the prescription from \citet{Madau_2015} and includes contributions from both hydrogen and helium reionization. Since $\tau$ is sensitive to the timing of reionization, lower (higher) values of $\tau$ indicate a later (earlier) onset of ionization.
%-------------------------------------------------%
\section{Methodology}
%-------------------------------------------------%
\subsection{Model Variants}
\label{subsec:model_var}
Our goal is to assess the viability of our source model in resolving the possible tension between low-redshift reionization and high-redshift JWST constraints. {In addition, we aim to test the constraining power of an integrated quantity, such as the UV luminosity density, $\rho_{\rm UV}$, against a non-integrated quantity}, such as the UV luminosity functions, $\phi_{\rm UV}$. To achieve these goals, we consider the following model variations.

\begin{enumerate}
    \item \textbf{{\eor}: } our fiducial reionization model, calibrated using EoR observables ($\dot{N}_{\rm ion} + x_{\rm HI} + \tau$), as described in \citet{2023ApJ...959....2B}. Here, we vary only the parameters $f_{\rm esc}$ and $C$, while keeping other parameters ({$A, B,$ and $D$}) fixed to the values obtained by fitting to the radiative transfer simulations of \citet{2018MNRAS.480.2628F}, to reduce the degeneracy among the parameters.  However, in this work, we update the set of observational constraints included in the likelihood to incorporate more recent observations. As previously shown in~\citet{2023ApJ...959....2B}, this calibration to EoR constraints uses a lower value of $D=2.28$, which serves as a proxy for weaker feedback (i.e., slower redshift evolution).

    \item \textbf{{\eorphiwf}:} the model jointly calibrated using reionization observables and the JWST UV luminosity functions ($\phi_{\rm UV} + \dot{N}_{\rm ion} + x_{\rm HI} + \tau$). Similar to \eor, this model adopts a fixed low value of $D = 2.28$, indicating weak feedback (wf), {and vary $f_{\rm esc}$ and $C$ while keeping $A, B,$ fixed along with $D$.}

     \item \textbf{{\eorphisf}:} similar to \textbf{{\eorphiwf}}, but varying all parameters ({$f_{\rm esc}, A, C$ and $D$}) except the $B$ parameter. This calibration, as will be seen later, favors models with a higher $D$ value by a factor $\sim \times 2$ as compared to other models, indicating stronger redshift evolution, which serves as a proxy for strong feedback (sf).
        
    \item \textbf{{\eorrhowf}:} the model combining reionization observables with the integrated UV luminosity density ($\rho_{\rm UV} + \dot{N}_{\rm ion} + x_{\rm HI} + \tau$). Similar to \eor, this model also adopts a fixed low value of $D = 2.28$, indicating weak feedback (wf), {and vary $f_{\rm esc}$ and $C$ while keeping $A, B,$ fixed along with $D$.}

     \item \textbf{{\eorrhosf}:} similar to \textbf{{\eorrhowf}}, but varying all parameters ({$f_{\rm esc}, A, C$ and $D$}) except the $B$ parameter. This calibration also favors models with a higher $D$ value by a factor $\sim \times 2$ as compared to other models, indicating stronger feedback (sf).
\end{enumerate}
It is worth noting that the primary motivation for fixing the amplitude (A), quenching mass scale (B), and feedback strength (D) in the {\bf EoR} and weak-feedback models ({\eorrhowf} $\&$ {\eorphiwf}) is that these parameters were previously found by \citet{2023ApJ...959....2B} and \citet{2016MNRAS.457.1550H} to successfully reproduce key reionization constraints. This approach, in turn, provides a well-defined benchmark for exploring variations of all parameters, which, as will be shown later, lead to the strong-feedback models ({\eorrhosf} $\&$ {\eorphisf}).

\begin{deluxetable*}{lcccc}
% \tablewidth{0.95\linewidth}
\tablecaption{Parameter priors used in our Bayesian analysis.\label{tab:priors}}
\tabletypesize{\footnotesize}
\tablehead{
\colhead{Model} &
\colhead{Parameter} &
\colhead{Prior} &
\colhead{Maximum} &
\colhead{Minimum}
} 
\startdata
All models & $f_{\rm esc}$ & uniform   & 0 & 1 \\
Strong feedback models & A & log-uniform   & 30 & 45 \\
All models & C & uniform   & -5 & 10 \\
Strong feedback models & D & uniform       & -5 & 10
\enddata
\tablecomments{The parameter B is initially varied for $\phi_{\rm UV}$ and $\rho_{\rm UV}$ model as described in appendix~\ref{appen:uvlf_rho}, then fixed to $\rm log_{10} (B/M_{\odot})=7.67$ to reduce the degeneracy among the parameters.} 
\end{deluxetable*}
%
%%%%%%%%%%%%%%%%%%%%%%%%%%%%%%%%%%%%%%%%%%%%%%%%%%%%%%%%%%%
\subsection{Bayesian Inference} \label{sec:infer}
%%%%%%%%%%%%%%%%%%%%%%%%%%%%%%%%%%%%%%%%%%%%%%%%%%%%%%%%%%%
We use a Bayesian inference framework to constrain the free parameters of our model by comparing its predictions with the observed data. The posterior probability distribution of the model parameters ${\theta}$ given the data $\mathcal{D}$ is calculated using Bayes’ theorem,
\begin{equation}
    \mathcal{P}({\theta}|\mathcal{D}) = \frac{\mathcal{L}(\mathcal{D}|{\theta}) \, \pi({\theta})}{P(\mathcal{D})},
\end{equation}
where $\mathcal{L}(\mathcal{D}|{\theta})$ is the likelihood function, $\pi({\theta})$ is the prior probability distribution on the parameters, and $\mathcal{P}(\mathcal{D})$ is the evidence. Since we are interested in sampling the posterior up to a normalization constant, the evidence $\mathcal{P}(\mathcal{D})$ is ignored in our analysis.

Assuming the datasets are statistically independent, the total likelihood is given by the product over the individual likelihoods,
\begin{equation}
    \mathcal{L}(\mathcal{D}|{\theta}) = \prod_{n} \mathcal{L}(\mathcal{D}_n|{\theta}),
\end{equation}
where $\mathcal{D}_n$ represents each observational dataset. More specifically,
\begin{enumerate}
    \item {\bf EoR}: $\mathcal{D}_n \subseteq \left\{x_{\rm HI}(z), \dot N_{\rm ion}(z), \tau \right\}$,
    \item \textbf{EoR-$\phi_{\rm UV}$-wf/sf}: $\mathcal{D}_n \subseteq \left\{\phi_{\rm{UV}}, x_{\rm HI}(z), \dot N_{\rm ion}(z), \tau \right\}$,
    \item  \textbf{EoR-$\rho_{\rm UV}$-wf/sf}:  $\mathcal{D}_n \subseteq \left\{\rho_{\rm{UV}}, x_{\rm HI}(z), \dot N_{\rm ion}(z), \tau\right\}$,
\end{enumerate}
where each term in the product takes the Gaussian form,
\begin{equation}
    \ln \mathcal{L}(\mathcal{D}_n|{\theta}) = -\frac{1}{2} \sum_i \left( \frac{\mathcal{D}_{n,i} - M_{n,i}({\theta})}{\sigma_{n,i}} \right)^2 ,
\end{equation}
where $\mathcal{D}_{n,i}$ and $\sigma_{n,i}$ are the observed data and uncertainties, and $M_{n,i}({\theta})$ are the model predictions for a parameter set ${\theta}$.

To sample the posterior distribution $\mathcal{P}({\theta}|D)$, we use the affine invariant Markov chain Monte Carlo (MCMC) ensemble sampler implemented in the publicly available python package {\it emcee} \citep{2013PASP..125..306F}. This method is particularly well-suited for high-dimensional, correlated parameter spaces, and enables efficient exploration of the posterior distribution. The resulting MCMC chains allow us to marginalize over nuisance parameters and derive credible intervals for each parameter. We adopt uniform or log-uniform priors depending on whether the parameter has a well-motivated range or spans multiple orders of magnitude, respectively. We consider the chains to have converged once the integrated autocorrelation time is well estimated.
Priors on the free parameters for different models are given in Table.~\ref{tab:priors}, motivated by the parameter values found in \citet{2023ApJ...959....2B} and \citet{2016MNRAS.457.1550H}.
\begin{figure}
    \centering
    \includegraphics[width=0.99\linewidth]{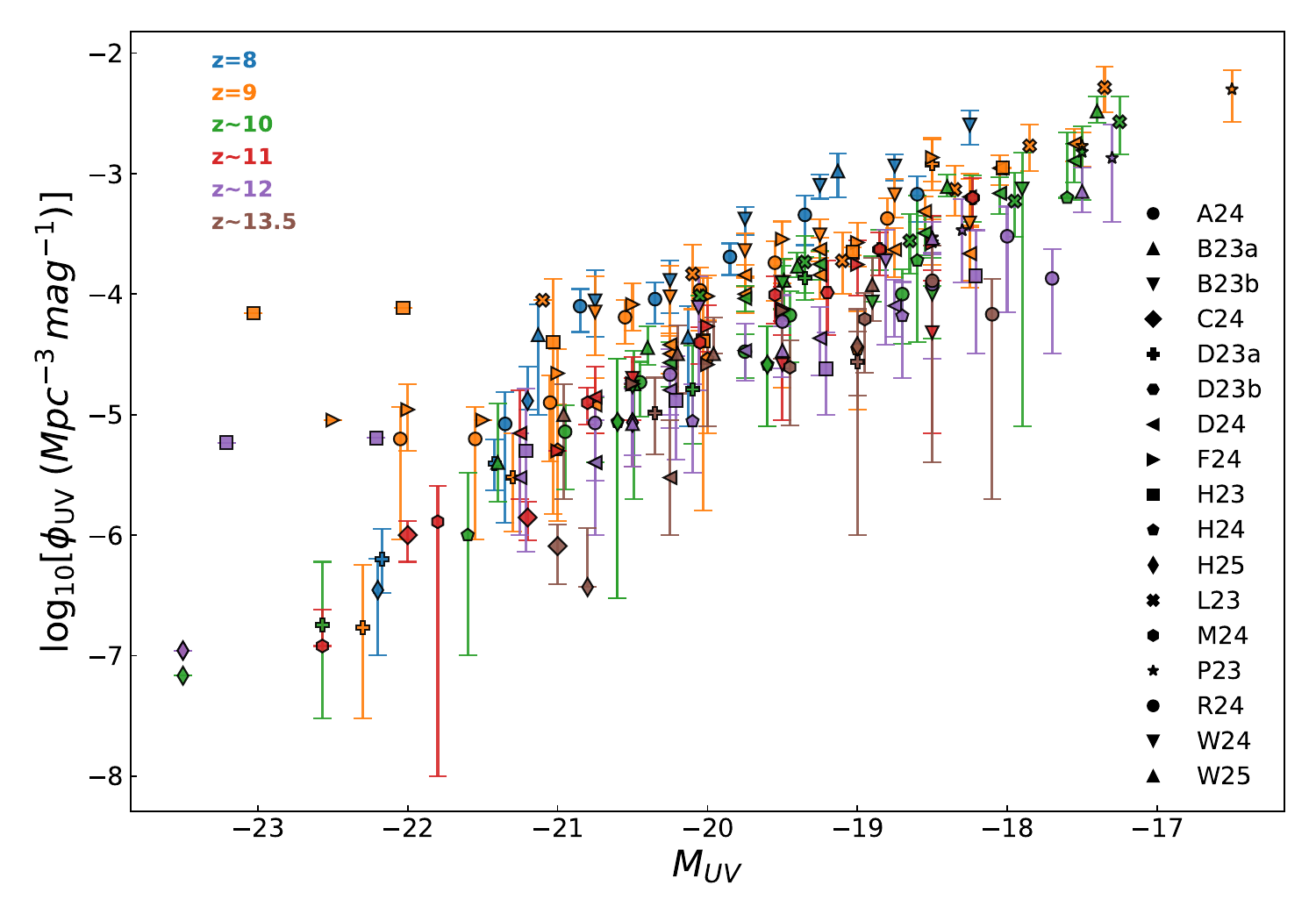}
    \caption{JWST UV Luminosity functions, $\phi_{\rm UV}$ observational data considered to constrain our models. Data points include~\citet[][A24]{2024ApJ...965..169A}, \citet[][B23a]{2023MNRAS.523.1009B}, \citet[][B23b]{2023MNRAS.523.1036B}, \citet[][C24]{2024ApJ...965...98C}, \citet[][D23a]{2023MNRAS.518.6011D}, \citet[][D23b]{2023MNRAS.520.4554D}, \citet[][D24]{2024MNRAS.533.3222D}, \citet[][F24]{2024ApJ...969L...2F}, \citet[][H23]{2023ApJS..265....5H}, \citet[][L23]{2023ApJ...954L..46L}, \citet[][P23]{2023ApJ...951L...1P}, \citet[][M24]{2024MNRAS.527.5004M}, \citet[][R24]{2024ApJ...970...31R}, \citet[][W24]{2024ApJ...966...74W}, \citet[][H24]{2024ApJ...960...56H}, \citet[][H25]{2025ApJ...980..138H}, \citet[][W25]{2025ApJ...992...63W}. Different colors represent different redshift bins, for example, $z=8$ (blue), $z=9$ (orange), $z\sim10 \, [9.8$--$10.75]$ (green), $z\sim11 \, [11$--$11.75]$ (red), $z\sim12 \, [12$--$12.8]$ (violet), and $z\sim13.5 \, [13$--$14.5]$ (brown). Different symbols denote data from individual studies.}
    \label{fig:uvlf_data}
\end{figure}
%
 %%%%%%%%%%%%%%%%%%%%%%%%%%%%%%%%%%%%%%%%%%%%%%%%%%%%%%%%%%%
\subsection{Observational constraints} \label{sec:obs}
%%%%%%%%%%%%%%%%%%%%%%%%%%%%%%%%%%%%%%%%%%%%%%%%%%%%%%%%%%%%
%
In this section, we present the observational constraints employed to calibrate and validate our models. We compile all the observational data using publicly available package {\it CoReCon} \citep{2023JOSS....8.5407G}.

\subsubsection{JWST UV Luminosity Function, \texorpdfstring{$\phi_{\rm UV}$}{phi UV}} \label{subsubsec:data_uvlf}
For the UV luminosity functions, we consider all current observational constraints from the following works ~\citet{2023ApJ...951L...1P, 2024ApJ...965..169A, 2023MNRAS.523.1009B, 2023MNRAS.523.1036B, 2024ApJ...965...98C, 2023MNRAS.518.6011D, 2023MNRAS.520.4554D, 2024MNRAS.533.3222D, 2024ApJ...969L...2F, 2023ApJS..265....5H, 2023ApJ...954L..46L, 2024MNRAS.527.5004M, 2024ApJ...970...31R, 2024ApJ...966...74W, 2025ApJ...992...63W}, spanning a redshift range from $z\, \sim\, 8 - 14$. Figure~\ref{fig:uvlf_data} shows these $\phi_{\rm UV}$ measurements, where different colors represent the observational data at different redshift bins and different symbols denote data from individual studies (e.g., blue circles: \citet{2024ApJ...965..169A} at $z\sim 8$, orange upper triangles: \citet{2023MNRAS.523.1009B} at $z\sim 9$, green plus: \citet{2023MNRAS.520.4554D} at $z\sim10$, purple squares: \citet{2023ApJS..265....5H} at $z\sim 12$, etc.).
\begin{deluxetable}{lcc}
\tablewidth{0.5\textwidth}
\tablecaption{UV luminosity densities ($\rho_{\rm UV}$) from recent JWST measurements used in this work.\label{tab:rhouv_obs}}
\tabletypesize{\footnotesize}
\tablehead{
\colhead{Redshift} &
\colhead{$\log(\rho_{\rm UV})$} &
\colhead{Reference}}
\startdata
11.0  & $25.15^{+0.13}_{-0.14}$ & \citet{2024MNRAS.527.5004M} \\
13.5  & $24.54^{+0.16}_{-0.25}$ & \\
\hline
9.0   & $25.40^{+0.10}_{-0.10}$ & \citet{2024ApJ...969L...2F} \\
11.0  & $25.00^{+0.20}_{-0.20}$ & \\
14.0  & $24.90^{+1.10}_{-0.70}$ & \\
\hline
9.0   & $25.17^{+0.18}_{-0.00}$ & \citet{2023ApJ...951L...1P} \\
10.75 & $25.12^{+0.12}_{-0.11}$ & \\
12.25 & $24.69^{+0.21}_{-0.18}$ & \\
\hline
8.0   & $25.80^{+0.04}_{-0.04}$ & \citet{2024ApJ...966...74W} \\
9.0   & $25.49^{+0.07}_{-0.07}$ & \\
10.25 & $24.90^{+0.12}_{-0.14}$ & \\
11.75 & $24.40^{+0.24}_{-0.29}$ & \\
\hline
9.0   & $25.28^{+0.19}_{-0.16}$ & \citet{2023ApJS..265....5H}\\
12.0  & $24.61^{+0.26}_{-0.26}$ & \\
\hline
8.0   & $25.61^{+0.00}_{-0.00}$ & \citet{2024ApJ...965..169A} \\
9.0   & $25.09^{+0.27}_{-0.24}$ & \\
10.5  & $24.84^{+0.50}_{-0.46}$ & \\
12.5  & $23.89^{+0.54}_{-1.45}$ & \\
\hline
9.0 &   $25.00^{+0.23}_{-0.27}$ & \citet{2024ApJ...960...56H} \\
10.0 &      $24.56^{+0.38}_{-0.39}$ & \\
12.0 &      $24.35^{+0.23}_{-0.31}$ & \\
\enddata
\tablecomments{All $\rho_{\rm UV}$ values are in units of erg\,s$^{-1}$\,Hz$^{-1}$\,Mpc$^{-3}$, integrated up to $M_{\rm UV} = -17$.}
\end{deluxetable}
\begin{deluxetable}{lcc}
\tablewidth{0.5\textwidth}
\tablecaption{Ionizing emissivity ($\dot{N}_{\mathrm{ion}}$) measurements.\label{tab:nion_obs}}
\tabletypesize{\footnotesize}
\tablehead{
\colhead{Redshift} &
\colhead{$\log(\dot{N}_{\mathrm{ion}})$} &
\colhead{Reference}}
\startdata
4.0  & $-0.14^{+0.80}_{-0.24}$ & \citet{2013MNRAS.436.1023B} \\
4.75 & $-0.01^{+0.80}_{-0.24}$ & \\
\hline
5.1  & $0.00^{+0.19}_{-0.14}$ & \citet{2021MNRAS.508.1853B} \\
6.0  & $0.54^{+0.73}_{-0.22}$ & \\
\hline
4.9  & $-0.22^{+0.22}_{-0.18}$ & \citet{2023MNRAS.525.4093G} \\
5.0  & $-0.25^{+0.27}_{-0.16}$ & \\
5.1  & $-0.21^{+0.34}_{-0.17}$ & \\
5.2  & $-0.18^{+0.30}_{-0.15}$ & \\
5.3  & $-0.20^{+0.29}_{-0.14}$ & \\
5.4  & $-0.19^{+0.31}_{-0.15}$ & \\
5.5  & $-0.23^{+0.31}_{-0.11}$ & \\
5.6  & $-0.18^{+0.26}_{-0.14}$ & \\
5.7  & $-0.20^{+0.24}_{-0.10}$ & \\
5.8  & $-0.22^{+0.30}_{-0.10}$ & \\
5.9  & $-0.20^{+0.23}_{-0.14}$ & \\
6.0  & $-0.15^{+0.22}_{-0.12}$ & \\
\enddata
\tablecomments{All $\dot{N}_{\mathrm{ion}}$ values are in units of $10^{51}$\,s$^{-1}$\,Mpc$^{-3}$.}
\end{deluxetable}
%-------------------------------------------------%
\subsubsection{JWST UV luminosity density, \texorpdfstring{$\rho_{\rm UV}$}{rho UV}} \label{subsubsec:data_rhouv}
We here consider the observational constraints reported in~\citet{2024ApJ...965..169A, 2024ApJ...969L...2F, 2023ApJS..265....5H, 2024MNRAS.527.5004M, 2023ApJ...951L...1P, 2024ApJ...966...74W}. All data with the 1-$\sigma$ errors are given in Table~\ref{tab:rhouv_obs}. We include $\rho_{\rm UV}$ measurements only at $z > 8$ in our likelihood analysis to ensure consistency with the post-JWST UVLF datasets considered in this work.
%
%-------------------------------------------------%
\subsubsection{Ionizing emissivity, \texorpdfstring{$\dot{N}_{\rm ion}$}{N ion}} \label{subsubsec:data_nion}  
We update the observational constraints on global ionizing photon emissivity over the redshift range $2 < z < 6$ used in \citet{2023ApJ...959....2B}. We incorporate the measurements from \citet{2021MNRAS.508.1853B, 2023MNRAS.525.4093G} along with \citet{2013MNRAS.436.1023B}, as shown in Table~\ref{tab:nion_obs}. {These emissivity measurements are derived from high-redshift QSO absorption spectra, where the hydrogen photoionization rate $\Gamma_{\rm HI}$ is inferred from the observed Ly-$\alpha$ forest opacity and mean free path of ionizing photons, which is then converted to an ionizing emissivity.}
%
%-------------------------------------------------%
\subsubsection{Neutral hydrogen fraction, \texorpdfstring{$x_{\rm HI}$}{x HI}} \label{subsubsec:data_xHI}
\begin{deluxetable}{cccc}
\tablewidth{0.4\textwidth}
\tabletypesize{\scriptsize}
\label{table:xHI_obs}
\tablecaption{IGM neutral hydrogen fraction, $x_{\rm HI}$ measurements}
%\tablewidth{0pt}
\tablehead{
\colhead{Redshift} & \colhead{Constraints} & \colhead{Observables} & \colhead{References}
}
\startdata
5.03 & $5.5^{+1.65}_{-1.42} \times 10^{-5}$ &  & \citet{2006AJ....132..117F} \\
5.25 & $6.7^{+2.44}_{-2.07} \times 10^{-5}$ & QSO &  \\
5.45 & $6.6^{+3.01}_{-2.47} \times 10^{-5}$ & absorption &  \\
5.65 & $8.8^{+4.6}_{-3.65} \times 10^{-5}$ & spectra &  \\
5.85 & $13^{+4.9}_{-4.08} \times 10^{-5}$ &  &  \\
6.1 & $4.3^{+3.0}_{-3.0} \times 10^{-4}$ &  &  \\ 
\hline
7.0 & $0.52^{+0.16}_{-0.16}$ & LAEs & \citet{2008ApJ...677...12O} \\ 
\hline
7.085 & $0.9^{+0.1}_{-0.1}$ & Ly$\alpha$ transmission profile & \citet{2011MNRAS.416L..70B} \\ 
\hline
7.0 & $0.75^{+0.15}_{-0.15}$ & LAEs & \citet{2012ApJ...744...83O} \\ 
\hline
7.0 & $0.34^{+0.09}_{-0.12}$ & LBGs & \citet{2014ApJ...795...20S} \\ 
\hline
7.1 & $0.6^{+0.19}_{-0.21}$ & Ly$\alpha$ damping wing & \citet{2017MNRAS.466.4239G} \\ 
\hline
7.09 & $0.52^{+0.26}_{-0.26}$ & QSO damping wing & \citet{2018ApJ...864..142D} \\
7.54 & $0.4^{+0.23}_{-0.20}$ &  &  \\ 
\hline
7.0 & $0.41^{+0.15}_{-0.11}$ & Ly$\alpha$ EW & \citet{Mason_2018} \\ 
\hline
7.5 & $0.21^{+0.17}_{-0.19}$ & QSO damping wing & \citet{2019MNRAS.484.5094G} \\ 
\hline
7.6 & $0.88^{+0.05}_{-0.1}$ & LBGs & \citet{2019ApJ...878...12H} \\ 
\hline
7.6 & $0.49^{+0.19}_{-0.19}$ & Ly$\alpha$ EW & \citet{2020ApJ...904..144J} \\ 
\hline
7.5 & $0.39^{+0.22}_{-0.13}$ &  & \citet{2020ApJ...897L..14Y} \\
5.4 & $5.63^{+0.56}_{-1.23} \times 10^{-5}$ & QSO damping wing &  \\
5.6 & $7.6^{+1.62}_{-0.61} \times 10^{-5}$ &  &  \\
5.8 & $8.85^{+1.76}_{-1.28} \times 10^{-5}$ &  &  \\ 
\hline
5.0 & $2.44^{+0.2}_{-0.05} \times 10^{-5}$ &  & \citet{2022MNRAS.514...55B} \\ 
5.1 & $2.65^{+0.13}_{-0.07} \times 10^{-5}$ & Ly$\alpha$ forest &  \\
5.2 & $2.99^{+0.12}_{-0.08} \times 10^{-5}$ &  &  \\
5.3 & $2.99^{+0.47}_{-0.13} \times 10^{-5}$ &  &  \\ 
\hline
4.9 & $2.53^{+0.44}_{-0.59} \times 10^{-5}$ &  & \citet{2023MNRAS.525.4093G} \\
5.0 & $2.26^{+0.45}_{-0.84}\times 10^{-5}$ &  &  \\
5.1 & $2.66^{+0.55}_{-1.34}\times 10^{-5}$ & QSO &  \\
5.2 & $2.75^{+0.56}_{-0.81}\times 10^{-5}$ & absorption &  \\
5.3 & $0.00051^{+0.0004}_{-0.0008}$ & spectra &  \\
5.4 & $0.0035^{+0.0025}_{-0.015}$ &  &  \\
5.5 & $0.0072^{+0.0035}_{-0.027}$ &  &  \\
5.6 & $0.016^{+0.0083}_{-0.025}$ &  &  \\
5.7 & $0.055^{+0.034}_{-0.071}$ &  &  \\
5.8 & $0.094^{+0.064}_{-0.062}$ &  &  \\
5.9 & $0.13^{+0.07}_{-0.13}$ &  &  \\
6.0 & $0.17^{+0.093}_{-0.092}$ &  &  \\ 
\hline
8.0 & $0.62^{+0.15}_{-0.36}$ & Ly$\alpha$ EW & \citet{2024ApJ...967...28N} \\
11.0 & $0.93^{+0.04}_{-0.07}$ &  &  \\ 
\hline
5.67 & $0.19 \pm 0.07$ & Ly$\alpha$ forest & \citet{2024AandA...688L..26S} \\ 
\hline
7.12 & $0.53^{+0.18}_{-0.47}$	& Ly$\alpha$ damping wing & \citet{2024ApJ...971..124U} \\
7.44 & $0.65^{+0.27}_{-0.34}$	&     & \\
8.28 & $0.91^{+0.09}_{-0.22}$	&     & \\
10.28 & $0.92^{+0.08}_{-0.10}$	&    & \\
\enddata
\end{deluxetable}
In compiling the dataset for the measurements of the volume-averaged neutral hydrogen fraction, ${x}_{\rm HI}$, we considered only measurements and omitted upper or lower limits in order to minimize potential biases arising from poorly constrained results. Our extensive dataset include measurements from \citet{2006AJ....132..117F, 2008ApJ...677...12O, 2011MNRAS.416L..70B,  2017MNRAS.466.4239G, 2019MNRAS.484.5094G, 2018ApJ...864..142D, 2012ApJ...744...83O, Mason_2018, 2019ApJ...878...12H, 2020ApJ...904..144J, 2020ApJ...897L..14Y, 2020ApJ...904...26Y, 2021MNRAS.508.1853B, 2022MNRAS.514...55B, 2023MNRAS.525.4093G, 2024ApJ...967...28N, 2024AandA...688L..26S, 2024ApJ...971..124U} using quasar absorption spectra, Ly$\alpha$ emitters (LAEs), Lyman break galaxies (LBGs), Ly$\alpha$ equivalent width (EW), Ly$\alpha$ forest, and Ly$\alpha$ damping wing over the redshift range $5 < z < 11$. All the data used in this work are given in Table~\ref{table:xHI_obs}.
%
%-------------------------------------------------%
\subsubsection{Optical depth, \texorpdfstring{$\tau$}{tau}} \label{subsubsec:data_tau}
For the Thomson optical depth, we use the \citet{2020A&A...641A...6P} results, which is
$\tau = 0.054 \pm 0.007$
at 68\% confidence level. This improved precision comes from better measurements of large-scale polarization, significantly tightening the constraint.
%
%-------------------------------------------------%
%%%%%%%%%%%%%%%%%%%%%%%%%%%%%%%%%%%%%%%%%%%%%%%%%%%%%%%%%%%
\section{Models calibration} \label{sec:model_calib}
%%%%%%%%%%%%%%%%%%%%%%%%%%%%%%%%%%%%%%%%%%%%%%%%%%%%%%%%%%%
%
\begin{figure*}[ht]
    \centering
    \includegraphics[width=0.45\textwidth]{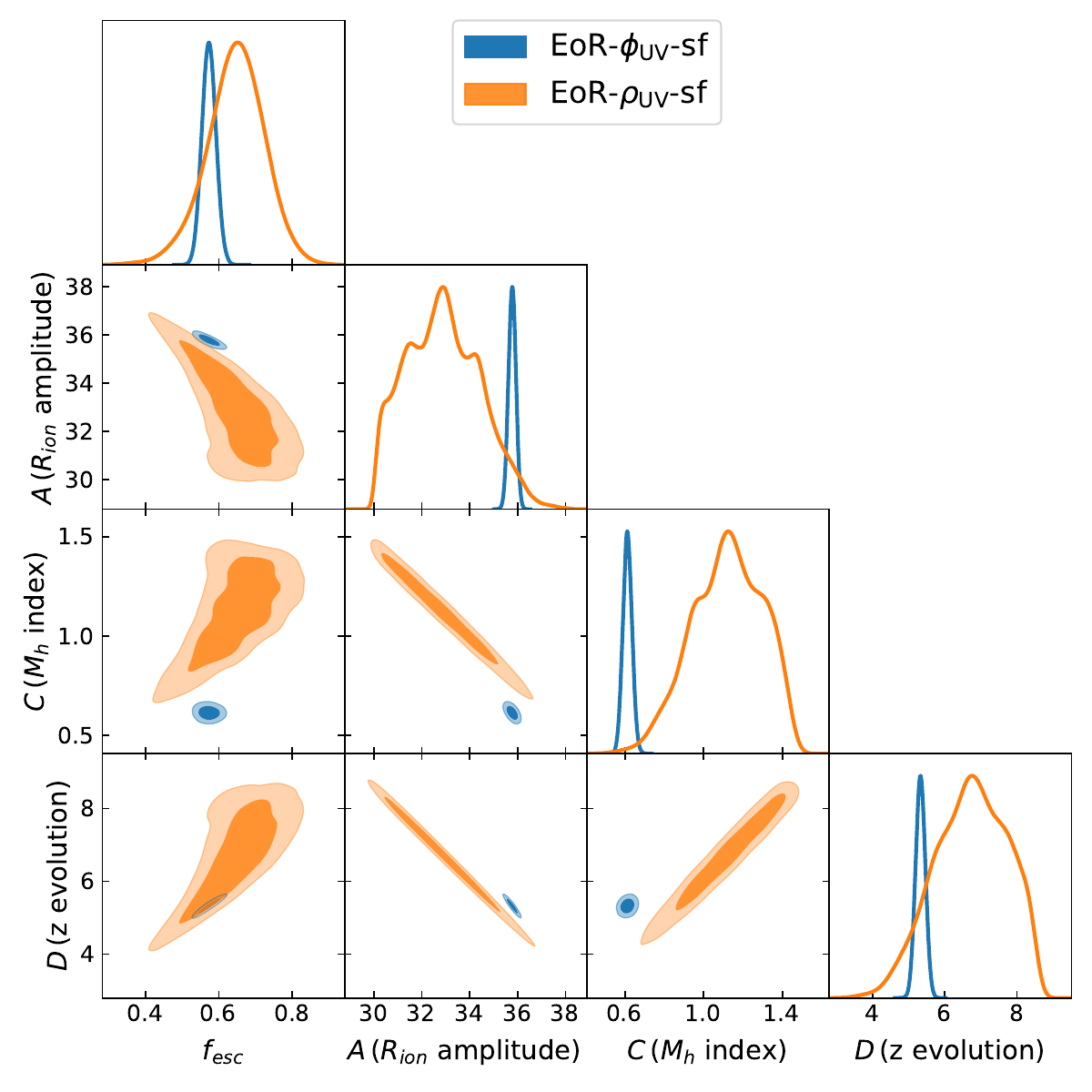}
    \includegraphics[width=0.45\textwidth]{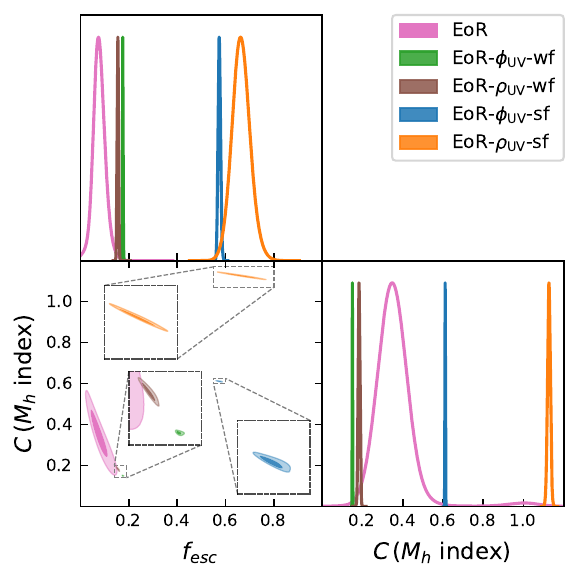}
    \caption{
    Corner plots showing the inferred parameter distributions for different combinations of observables:
    \textbf{Left:} Constraints from joint fits combining EoR observables with $\phi_{\rm UV}$ and $\rho_{\rm UV}$ data, fixing $\rm log_{10}(B/M_{\odot})= 7.67$, to reduce the parameter degeneracies as described in appendix~\ref{appen:uvlf_rho}.
    \textbf{Right:} Constraints from EoR and joint fits for all models described in \S\ref{subsec:model_var}, allowing only $f_{\rm esc}$ and $C$ to vary and fixing other parameters to the values given in Table.~\ref{tab:all_model}. 
    }
    \label{fig:combined_corner_plots}
\end{figure*}
We now consider the combined constraints from recent JWST observations and globally-averaged observational quantities of reionization described in \S\ref{subsubsec:data_uvlf}, \S\ref{subsubsec:data_rhouv}, \S\ref{subsubsec:data_nion}, \S\ref{subsubsec:data_xHI}, and \S\ref{subsubsec:data_tau} respectively, to calibrate all our models. The best-fit parameter values for all models are summarized in Table~\ref{tab:all_model}. 

We present this comparison in the left panel of Figure~\ref{fig:combined_corner_plots}. The parameters in the \eorphisf case yields significantly tighter constraints, as is evident from the narrower contours and sharper marginalized peaks (shown in {blue}). This is primarily because the UVLF retains magnitude-resolved information about galaxy populations, allowing a more precise mapping between halo mass and galaxy luminosity. 
However, other EoR observables (e.g. $\dot N_{\rm ion}, x_{\rm HI}$ and $\tau$) play a crucial role in constraining the allowed ionizing budget, resulting in a similar tighter constraint on $f_{\rm{esc}}$ in the case of the \eorphisf model. In the 2-D posterior distributions, we observe a negative correlation between the amplitude (A) and $f_{\rm esc}$, which arises naturally from the $R_{\rm ion}$ parameterization, as these parameters are multiplied by each other. Positive correlations are observed between $f_{\rm esc}$ and both the feedback strength (D) and the $M_{\rm h}$ index (C), indicating that a higher $f_{\rm esc}$ is required to counterbalance the increased feedback strength and the reduced contribution from low-mass halos. As seen in Table~\ref{tab:all_model}, by allowing the redshift evolution (D parameter i.e. feedback strength) to vary, we obtain a higher values of D $\sim$ 6.8 for the \eorrhosf and D $\sim$ 5.3 for the \eorphisf models, which is approximately $\sim 2-3 \times$ higher as compared to the fiducial value D $\sim$ 2 that is adopted in our previous analysis in \citet{2023ApJ...959....2B, 2016MNRAS.457.1550H} and in other models such as \eor, \eorphiwf and \eorrhowf. 

We now turn our attention to the right panel of Figure~\ref{fig:combined_corner_plots}. Here we present a comparison for the posterior distributions for only the parameters of $f_{\rm esc}$ and the mass index $C$. In this analysis, we follow our previous approach in \citet{2023ApJ...959....2B, 2016MNRAS.457.1550H} for {\eor}, {\eorphiwf} and {\eorrhowf}, where we fix the parameters $\rm log_{10}(A/M_{\odot}^{-1}\, s^{-1}) = 40, \rm log_{10} (B/M_{\odot}) = 7.67, D = 2.28$. In {\eorphisf} and {\eorrhosf} cases, we fix the parameters to their respective values given in the last two rows in Table.~\ref{tab:all_model}, and vary only $f_{\rm esc}$ and $C$ in all models. By focusing on the $f_{\rm esc}$-$C$ plane, we can determine the contribution of different galaxy populations responsible to driving different observables at different redshifts including reionization. 
For the {\eor} model, we find a consistent $f_{\rm{esc}} - C$ relation (in {{pink}}) with our previous findings with a slight increase in $f_{\rm{esc}} = 0.075_{-0.020}^{+0.025}$, and slight decrease in $C = 0.349_{-0.062}^{+0.062}$, as compared to $f_{\rm{esc}} = 0.02_{-0.01}^{+0.02}$, $C = 0.56_{-0.17}^{+0.19}$ found in \citet{2023ApJ...959....2B} due to adding the most updated measurements in the likelihood. In this figure, we also show the results for {\eorphiwf} (in {green}) and {\eorrhowf} (in {brown}) models, where $f_{\rm esc}$ increases to $0.174 \pm 0.001$ and $0.154 \pm 0.003$, and $C$ decreases to $0.15 \pm 0.001$ and $0.183_{-0.004}^{+0.005}$, respectively. This shows that adding UV observables, in presence of week feedback, prefers models with higher $f_{\rm esc}$ and more contribution from low mass halos. In the strong feedback cases, \eorphisf (in {blue}) and \eorrhosf (in {orange}), both $f_{\rm esc}$ and C parameters increase. For example, $f_{\rm esc}$ increases to $0.573 \pm 0.019$ and $0.648_{-0.079}^{+0.074}$, and $C$ increases to $0.612 \pm 0.023$ and $1.127_{-0.193}^{+0.188}$, respectively, preferring models with much higher $f_{\rm esc}$ and much more contribution from massive halos.   

\begin{figure*}[t]
    \centering
    \includegraphics[width=0.95\textwidth]{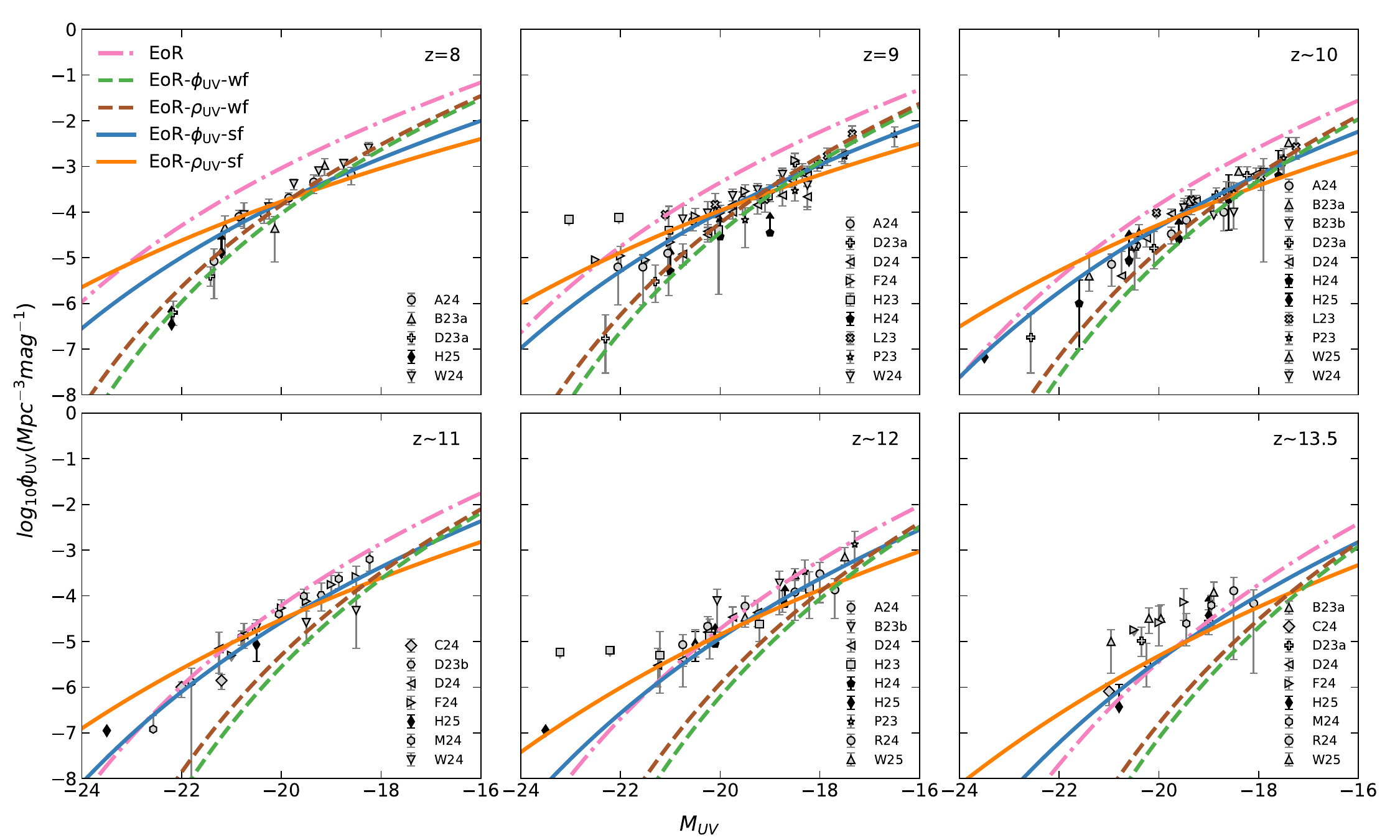}
    \caption{Comparison of Rest-frame ultraviolet luminosity functions, (UVLFs), as predicted by our models using inferred parameter values given in Table~\ref{tab:all_model} with the JWST observational constraints from ~\citet[][A24]{2024ApJ...965..169A}, \citet[][B23a]{2023MNRAS.523.1009B}, \citet[][B23b]{2023MNRAS.523.1036B}, \citet[][C24]{2024ApJ...965...98C}, \citet[][D23a]{2023MNRAS.518.6011D}, \citet[][D23b]{2023MNRAS.520.4554D}, \citet[][D24]{2024MNRAS.533.3222D}, \citet[][F24]{2024ApJ...969L...2F}, \citet[][H23]{2023ApJS..265....5H}, \citet[][L23]{2023ApJ...954L..46L}, \citet[][P23]{2023ApJ...951L...1P}, \citet[][M24]{2024MNRAS.527.5004M}, \citet[][R24]{2024ApJ...970...31R}, \citet[][W24]{2024ApJ...966...74W}, \citet[][H24]{2024ApJ...960...56H}, and \citet[][H25]{2025ApJ...980..138H}. The observational data points are same as Figure~\ref{fig:uvlf_data}. To avoid ambiguity with the lines, we use gray color for photometric samples and black for spectroscopically confirmed observations. Weak feedback models struggle to capture the elevated UVLF at $z>9$ (in dashed {brown} and {green}), while strong feedback models (in solid {blue} and {orange}) successfully reproduce the elevated UVLF at high-redshift ($z\gtrsim10$), but over-estimate the bright end at $z<9$.}
    \label{fig:uvlf}
\end{figure*}
\begin{deluxetable*}{lccccc}
\label{tab:all_model}
%\tabletypesize{\scriptsize}
\tablecaption{Best-fit parameters.}
\tablehead{
    \colhead{Model} & \colhead{$\rm f_{\rm esc}$} & \colhead{$\rm log_{10}(A/M_{\odot}^{-1}\, s^{-1})$} & \colhead{$\rm C$} & \colhead{$\rm D$} & \colhead{Constraints}
}
\startdata
\\
EoR     & $0.075^{+0.025}_{-0.020}$ & $40.00$ [fixed] & $0.349 \pm 0.062$ & $2.28$ [fixed] & $\dot N_{ion} + x_{\rm HI} + \tau$ \\[1ex]
{\eorphiwf} & $0.174 \pm 0.001$ & $40.00$ [fixed] & $0.15 \pm 0.001$ & $2.28$ [fixed] & $\dot N_{ion} + x_{\rm HI} + \tau + \phi_{\rm UV}$\\[1ex]
{\eorrhowf} & $0.154 \pm 0.003$ & $40.00$ [fixed] & $0.183^{+0.005}_{-0.004}$ & $2.28$ [fixed] & $\dot N_{ion} + x_{\rm HI} + \tau + \rho_{\rm UV}$ \\[1ex]
{\eorphisf} & $0.573 \pm 0.019$ & $35.771_{-0.156}^{+0.153}$ & $0.612 \pm 0.023$ & $5.327 \pm 0.133$ & $\dot N_{ion} + x_{\rm HI} + \tau + \phi_{\rm UV}$\\[1ex]
{\eorrhosf} & $0.648^{+0.074}_{-0.079}$ & $32.829^{+1.657}_{-1.618}$ & $1.127^{+0.188}_{-0.193}$ & $6.767^{+1.052}_{-1.095}$ & $\dot N_{ion} + x_{\rm HI} + \tau + \rho_{\rm UV}$ \\[1ex]
\enddata
\tablecomments{The parameter B is fixed to $\rm log_{10} (B/M_{\odot})=7.67$.}
\end{deluxetable*}
%\renewcommand{\arraystretch}{1.0} % reset to default
%
%-------------------------------------------------%
%%%%%%%%%%%%%%%%%%%%%%%%%%%%%%%%%%%%%%%%%%%%%%%%%%%%%%%%%%%
\section{Results} \label{sec:res}
%%%%%%%%%%%%%%%%%%%%%%%%%%%%%%%%%%%%%%%%%%%%%%%%%%%%%%%%%%%
Here we assess the ability of our models to reproduce various observables, and alleviate the potential tension between JWST and EoR constraints. 

\subsection{The UV luminosity function, \texorpdfstring{$\phi_{\rm UV}$}{phi UV}}
\label{res:uvlf_res}
Using the best-fit values shown in Table~\ref{tab:all_model}, we compare all models to the comprehensive set of UVLF constraints from recent JWST surveys across a wide redshift range, $z \sim 8$ to $z \sim 14$, shown in Figure~\ref{fig:uvlf}. Each panel illustrates the models evaluated at a specific redshift bin, alongside observational data points from multiple surveys including CEERS, GLASS, JADES, COSMOS-Web, PEARLS, and NGDEEP \citep{2022ApJ...935..110T, 2023arXiv230602465E, 2024ApJ...965...98C, 2024ApJ...969L...2F, 2023AJ....165...13W, 2024ApJ...965L...6B}. The dot-dashed curve in {pink}, the dashed curves in {green}, and {brown}; and the solid curves in {blue}, and {orange} in each panels show the result of model predictions for {\bf EoR}, {\eorphiwf}, {\eorrhowf}, {\eorphisf}, and {\eorrhosf} respectively. The gray symbols in Figure~\ref{fig:uvlf} are estimates based on photometric samples and black symbols represent the number densities of galaxies with spectroscopic redshifts.

The {\bf EoR} model over-predicts the abundance of galaxies at all magnitudes up to $z \sim 10$ due to the high overall amplitude (higher A value), and marginally under-predicts the bright-end at higher redshifts $z > 10$, which may be expected since this model is not calibrated to any JWST $\phi_{\rm UV}$ or $\rho_{\rm UV}$ observations.  
All the weak-feedback model predictions from {\eorphiwf} and {\eorrhowf} tend to fit the observations well compared to other models up to $z\sim10$ but under-predict the number density of bright galaxies at redshifts $z>10$, as the power-law slope is a few times lower and the redshift evolution/feedback is mild compared to the other cases. 

The {\eorphisf} model shows excellent agreement at the faint end with the observational UVLFs for the entire redshift range $z \sim 8$ to $z \sim 14$, reproducing both the shape and normalization of the luminosity function. However, this requires a more rapid redshift evolution, $R_{\rm ion} \propto (1+z)^{5.32}$ as compared to $(1+z)^{2.28}$ as derived from radiative transfer simulations \citep[][]{2017MNRAS.468..122H}.
The agreement at the faint end validates the model’s assumptions about halo mass-to-light ratios and suppression for low-mass halos due to feedback are robust. The better fit of {\eorphisf} model over {\eorphiwf} at the bright end ($M_{\rm UV} \leq -21$) at $z \geq 10$ is particularly notable, which suggests that a strong feedback from early massive galaxies is essential to explain the elevated UVLF observed in recent JWST surveys \citep[e.g.][]{2023MNRAS.523.1009B, 2023MNRAS.523.1036B, 2023MNRAS.518.6011D, 2024ApJ...965..169A} at high redshifts. 
The {\eorrhosf} model tends to under-predict the faint-end, $M_{\rm UV} > -20$ of the UV luminosity function. This is because $\rho_{\rm UV}$ is an integrated quantity and does not retain information about the shape of the luminosity function, resulting in a lower normalization ($\log_{10} A = 32.829^{+1.657}_{-1.618}$) and correspondingly higher values for the slope and redshift evolution parameters compared to the {\eorphisf} case.

%-------------------------------------------------%
\subsection{Globally averaged quantities} \label{res:global_quan}
We now proceed to present our model predictions for the globally averaged quantities, including $\rho_{\rm UV}$, $\rho_{\rm SFR}$, $x_{\rm HI}$, $\dot{N}_{\rm ion}$, and $\tau$, alongside the observational data, as shown in Figure~\ref{fig:all_global_cons}. 
\begin{figure*}[!htbp]
    \centering
    % ---------- Left column ----------
    \begin{minipage}[t]{0.46\textwidth}
        \centering
        \includegraphics[width=\linewidth, trim=2mm 0 0 0, clip]%
            {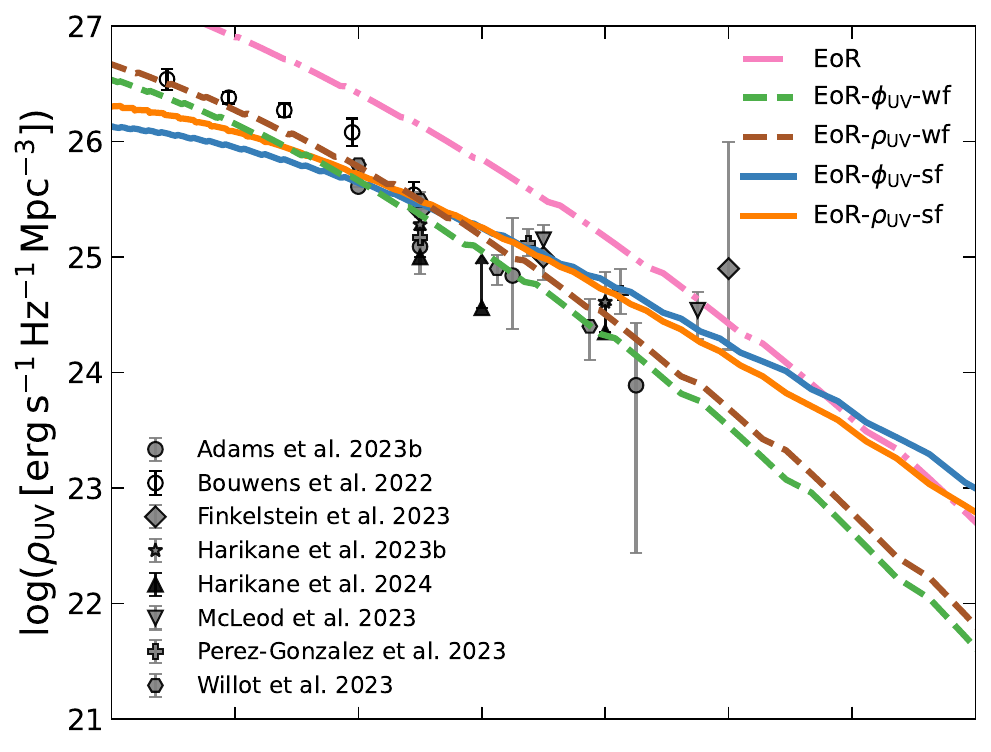}\\[1.5mm]
        \includegraphics[width=\linewidth, trim=0 0 3mm 0, clip]%
            {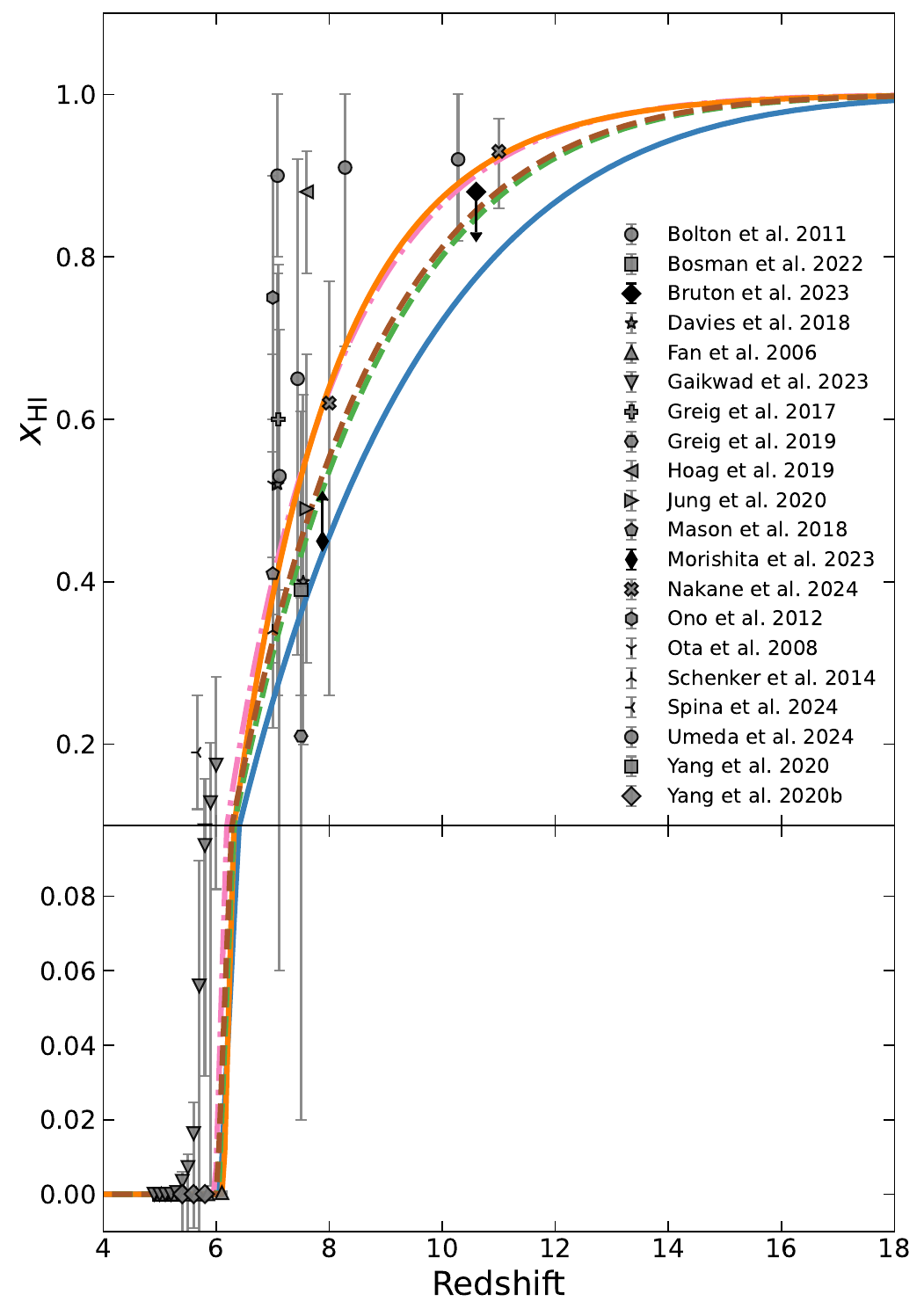}
    \end{minipage}
    \hfill
    % ---------- Right column ----------
    \begin{minipage}[t]{0.48\textwidth}
        \centering
        \includegraphics[width=\linewidth, trim=2mm 0 0 0, clip]%
            {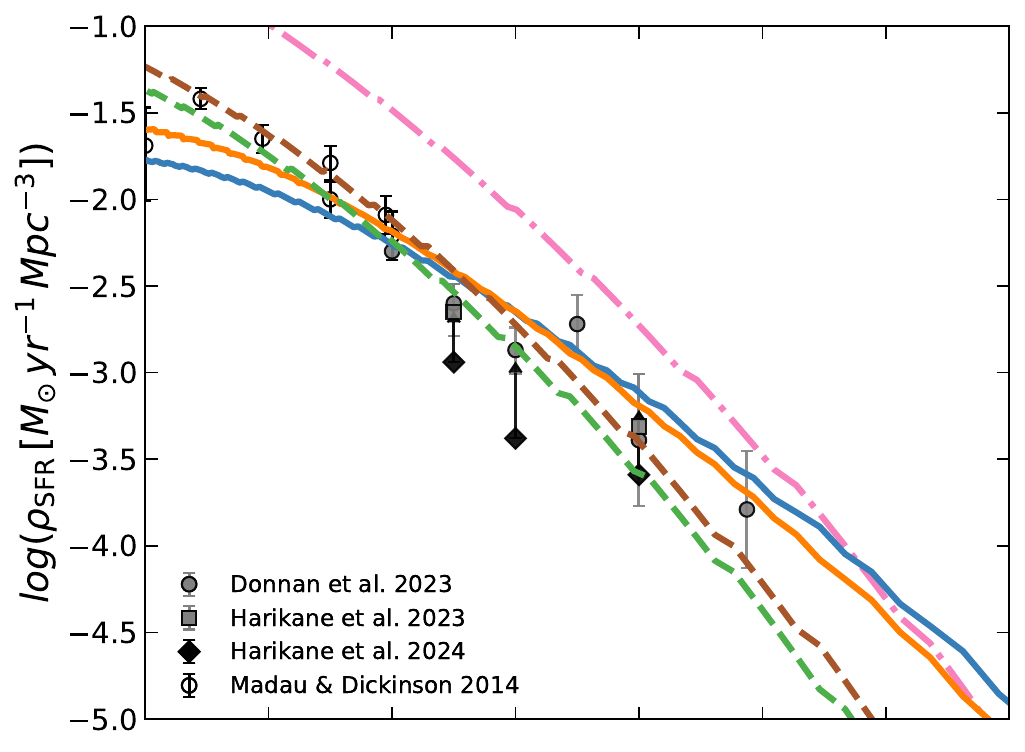}\\[1.5mm]
        \includegraphics[width=\linewidth, trim=2mm 0 0 0, clip]%
            {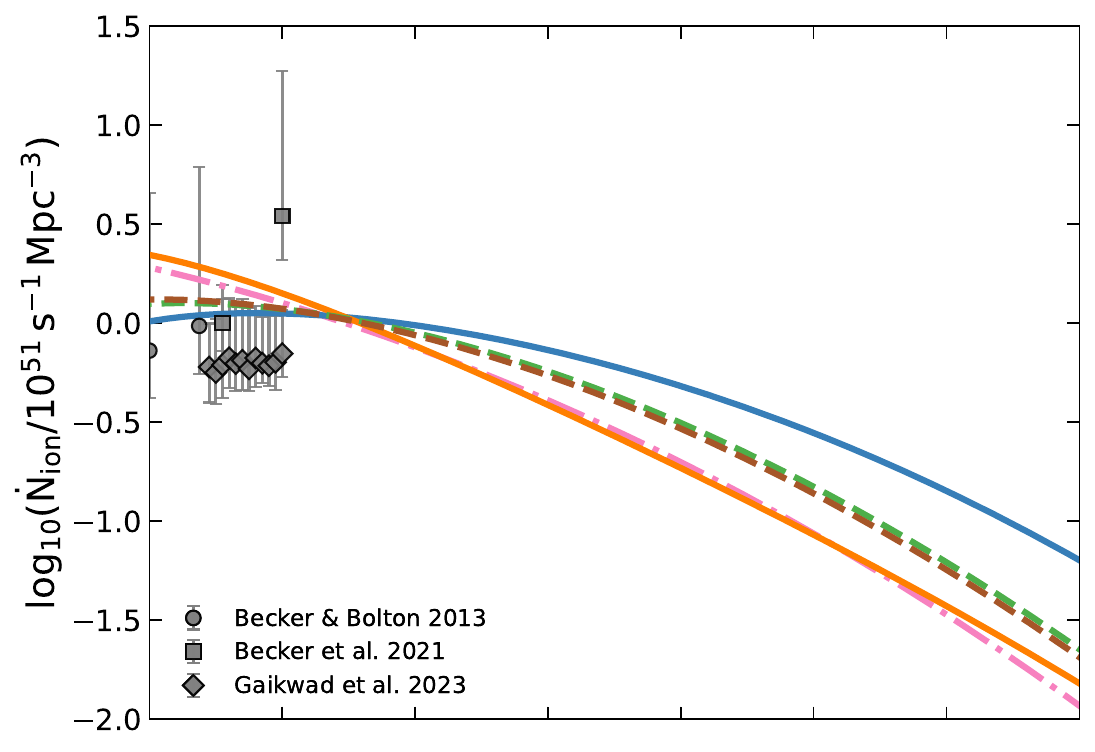}\\[1.5mm]
        \includegraphics[width=\linewidth, trim=0 0 3mm 0, clip]%
            {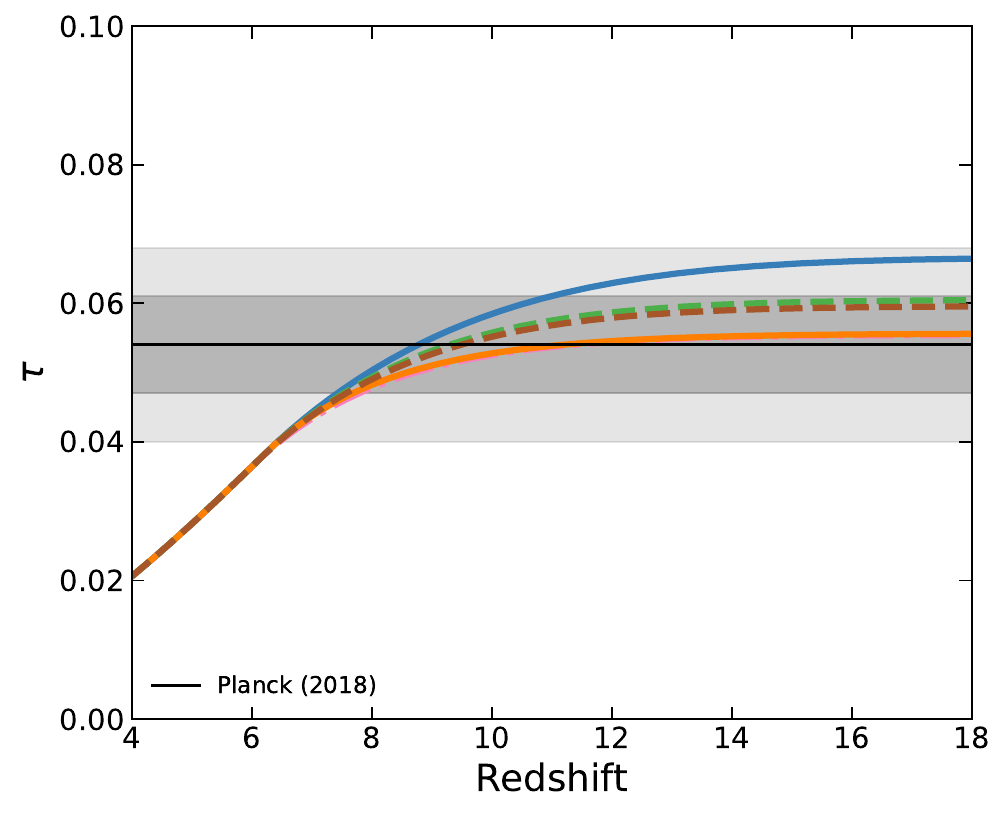}
    \end{minipage}
    \caption{Redshift evolution of all globally averaged quantities comparing with the observations given in \S\ref{sec:obs}. {\bf Top left:} UV luminosity density, $\rho_{\rm UV}$, {\bf top right:} star formation rate density, $\rho_{\rm SFR}$, {{\bf bottom left:}} neutral hydrogen fraction, $x_{\rm HI}$, {\bf middle right:} ionizing emissivity, $\dot N_{\rm ion}$, and {\bf bottom right:} Thomson optical depth, $\tau$ for all the model variants given in Table.~\ref{tab:all_model}. All models reproduce the reionization history within the 1–2 $\sigma$ level of current observational constraints, but their ability to match JWST observations varies significantly. Despite agreeing well with the $x_{\rm HI}$, and Planck optical depth measurement, the \textbf{EoR} model (in dot-dashed {pink}) fails to capture the evolution of the JWST UV luminosity density. 
    The weak-feedback models (in dashed {brown} and {green}), with higher $f_{\rm esc}$ values and a larger contribution from faint galaxies than the {\bf EoR} model, reionize earlier than it. In contrast, the {\eorphisf} model (in solid {blue}), featuring strong feedback and an enhanced contribution from massive galaxies, produces relatively high ionizing emissivity, consequently, an early yet more extended reionization history, still consistent with the upper and lower limits of \citet{2023ApJ...949L..40B} (black thick diamond) and \citet{2023ApJ...947L..24M} (black diamond) respectively {(bottom left panel)}.}
    \label{fig:all_global_cons}
\end{figure*}
\begin{figure}
    \centering
    \includegraphics[width=0.99\linewidth]{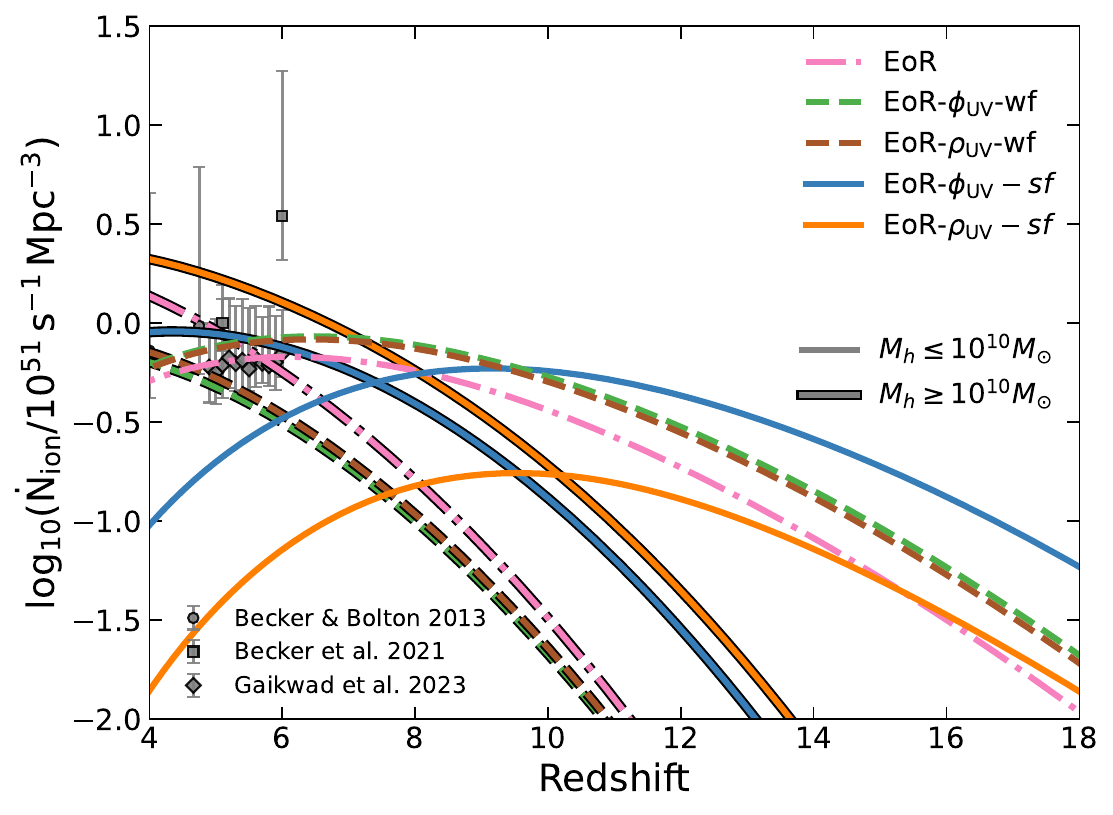}
    \caption{The ionizing emissivity for all models, separating the contribution of massive ($M_h \geq 10^{{10}} M_{{\odot}}$) and faint galaxies ($M_h \leq 10^{{10}} M_{{\odot}}$) by {black-outlined and no-outlined} curves respectively.}
    \label{fig:Rion}
\end{figure}
%
%-------------------------------------------------%
\subsubsection{UV luminosity density \& star formation rate density} \label{res:rho_res}
Using the best-fit parameters from Table.~\ref{tab:all_model}, we show the redshift evolution of $\rho_{\rm UV}$, and the predicted evolution of the cosmic star formation rate density, $\rho_{\rm SFR}$ compared to the observational data from recent JWST surveys 
in the top left and right panels of Figure~\ref{fig:all_global_cons} respectively.

{The EoR constraints alone lead} to an over-prediction of $\rho_{\rm UV}$, and $\rho_{\rm SFR}$ across all redshifts ($z \sim 8-14$). The addition of $\phi_{\rm UV}$ or $\rho_{\rm UV}$ constraints leads to an overall decrease in the redshift evolution of $\rho_{\rm UV}$, and consequently of $\rho_{\rm SFR}$ at $z<14$, improving agreement with the observations. 
While the models {\eorphiwf} and {\eorrhowf} improves agreement at $z < 10$, {and are broadly consistent} with expectations from weak feedback models, similar to \citet{2024ApJ...965..169A, 2024ApJ...966...74W}, they under-predict the enhanced $\rho_{\rm UV}$ at $z > 10$, {highlighting that a weak feedback model} may not fully capture the abundance of UV-bright galaxies revealed by JWST during the cosmic dawn. We also note that the observational uncertainties at $z > 10$ span several orders of magnitude, offering limited information on the precise evolution at high redshift. As a result, these measurements are insufficient to robustly constrain our models at $z>12$.

Both the {\eorphisf} and {\eorrhosf} results in better consistency with the observed $\rho_{\rm UV}$ \citep{2023ApJS..265....5H, 2024ApJ...969L...2F, 2024MNRAS.527.5004M, 2023ApJ...951L...1P}, and $\rho_{\rm SFR}$ \citep{2023MNRAS.518.6011D, 2023ApJS..265....5H}, particularly in the redshift range $z \sim 8$–$12$. However, \textbf{sf} models fail to capture the increase in the UV luminosity density, suggested by Hubble Frontier Fields (HFF) clusters observation \citet{2022ApJ...940...55B} (open square points in  top left panel of Figure~\ref{fig:all_global_cons}) at redshifts $z<7$. We verify that the inclusion of low-redshift $\rho_{\rm UV}$ data drives the model toward better consistency with the overall evolutionary trend across the full redshift range ($z = 2$–$16$). Similarly, at lower redshifts ($z < 7$), both \textbf{sf} models deviate from the empirical \citet{2014ARA&A..52..415M} fit (open round points in  top right panel of Figure~\ref{fig:all_global_cons}), underestimating the sharp rise in the star formation rate densities. {We also note that all our models predict $\rho_{\rm UV}$ and $\rho_{\rm SFR}$ values that lie above
the lower limits set by the spectroscopically confirmed measurements
of \citet{2024ApJ...960...56H}.}

%-------------------------------------------------%
\subsubsection{EoR observables} \label{res:eor_res}
We now show the redshift evolution of volume-averaged neutral hydrogen fraction, $x_{\rm HI}(z)$, and the model-predicted ionizing emissivity histories in {bottom left and middle right panels} of Figure~\ref{fig:all_global_cons} respectively. All the models demonstrate broad consistency with the observational constraints from \citet{2013MNRAS.436.1023B} over the range $z = 4$--$5.5$, where the emissivity reaches a few ionizing photons per hydrogen atom per Gyr.

The \textbf{EoR} and {\eorrhosf} models show a sharp declining emissivity with increasing redshift, resulting in a delayed onset of reionization, starting around $z \sim 12$, but shows a rapid decline in $x_{\rm{HI}}$ after $z \sim 7.5$, completing reionization by $z \sim 6$. For instance, between $z \sim 6.5$ and $z \sim 7.5$, these models remain consistent within the 1–2$\sigma$ bounds of damping wing constraints from quasars (\citealt{2017MNRAS.466.4239G}, \citealt{2018ApJ...864..142D}, \citealt{2020ApJ...897L..14Y}) and Ly$\alpha$ emitter fractions (\citealt{2012ApJ...744...83O}, \citealt{Mason_2018}, \citealt{2020ApJ...904...26Y}). At $z \sim 7.5$, these models yield $x_{\rm HI} \sim 0.4$--$0.5$, capturing the midpoint of reionization and indicating a partially neutral IGM. However, at earlier times ($z \gtrsim 8$), the neutral fraction rises steeply to $x_{\rm HI} \sim 0.7$--$0.9$, and these two models maintain the highest neutral fraction at $z > 7$, making it more consistent with recent JWST-informed constraints on a late and rapid reionization \citep{2024ApJ...967...28N, 2024ApJ...971..124U}. 
On the other hand, the ionizing emissivity of {\eorphisf} model lies systematically above the \textbf{EoR} and {\eorrhosf} fit, especially at higher redshifts ($z \gtrsim 8$), 
implying that enhanced emissivity naturally arises when calibrated to the elevated UVLF measured by JWST.
Consequently, the {\eorphisf} model shows a very early and extended reionization history, beginning around $z \sim 18$ and suggesting the universe is at least $20 \%$ ionized by $z\sim10$ as opposed to $< 10\%$ according to \citet{2020A&A...641A...6P}. However, this model is consistent with the JWST-informed lower limit of \citet{2023ApJ...947L..24M} (black triangle) and the upper limit of \citet{2023ApJ...949L..40B} (black thick triangle).
Additionally as this model exhibits a relatively high ionizing emissivity at early times, followed by a gradual decline toward $z \sim 6$, this leads to a reionization history that also completes around $z \sim 6$, in agreement with the other two models. In particular, At $z < 6$, all models converge toward a nearly ionized IGM with $x_{\rm HI} \sim 0.01$--$0.05$, in agreement with constraints from \citet{2006AJ....132..117F}, \citet{2020ApJ...904...26Y}, and \citet{2023MNRAS.525.4093G}.
The {\eorphiwf} and {\eorrhowf} models exhibit intermediate reionization histories as this model predicts higher emissivity than the \textbf{EoR}-only case but lower than the {\eorphisf} fit, at the high redshifts, $z>7$. 
Notably, none of the models favor the high emissivity of \citet{2021MNRAS.508.1853B} at $z\sim6$, instead, all the models align more closely with the findings of \citet{2013MNRAS.436.1023B} and \citet{2023MNRAS.525.4093G}, supporting a reionization completion around $z \sim 6$.

Despite the close relationship between the $\phi_{\rm UV}$ and $\rho_{\rm UV}$, the UVLF-constrained model still predicts a noticeably different reionization timeline. As upcoming JWST surveys provide more detailed information on high-redshift galaxies, our understanding of the reionization timeline is expected to improve significantly.  
In parallel, forthcoming 21-cm experiments, e.g., the Square Kilometre Array (SKA-low) will deliver the first spatially resolved 21-cm maps of the neutral intergalactic medium, directly tracing how ionized regions expand around early sources, and providing direct tests of reionization models. 

The cumulative electron scattering optical depth, shown in lower right panel of Figure~\ref{fig:all_global_cons}, reflects the integrated history of reionization. All models are consistent with the \citet{2020A&A...641A...6P}, being within the 2-$\sigma$ level. In particular, the \textbf{EoR} and {\eorrhosf} models produce the lowest $\tau$, while the {\eorphisf} fit yields slightly higher values, yet within 2-$\sigma$ of \citet{2020A&A...641A...6P} value due to early yet more extended reionization, which prolongs partial ionization epochs. The \textbf{wf} models are also within 1-$\sigma$ of \citet{2020A&A...641A...6P}. The differences remain modest, indicating that all the models are viable within current CMB constraints. 

To further verify the implied role of different galaxy populations in driving reionization from Table~5, we split the total ionizing emissivity, $\dot{N}_{\rm ion}$, into two components: emissivities from galaxies fainter and brighter than $10^{10} M_{\odot}$. We then present the evolution of these two components for all models in Figure~\ref{fig:Rion}, indicating the emissivities from galaxies fainter/brighter than $10^{10} M_{\odot}$ with {no-outlined/black-outlined} lines, respectively. As shown earlier in Figure~\ref{fig:all_global_cons}, reionization begins earlier in the {\bf wf} models compared to the {\rm EoR} model, due to the higher contributions to the emissivity budget from faint galaxies ({no-outlined green}/{brown} vs. {pink} lines), as seen in Figure~\ref{fig:Rion}. This is consistent with the lower inferred $C$ values by roughly a factor of 2 in these models compared to the {\bf EoR} model, as reported in Table~\ref{tab:all_model}. This table also shows that in the {\bf sf} models, the $C$ values are higher by a factor of 2--4 compared to the {\bf EoR} model, indicating a much stronger contribution from massive halos. Comparing the {black-outlined orange}/{blue} lines with the {pink} lines in Figure~\ref{fig:Rion}, we clearly see that the contribution from massive halos in the {\bf sf} models is higher than in the {\bf EoR} model. This figure vividly illustrates that a stronger contribution from bright galaxies, combined with more pronounced redshift evolution (i.e., stronger feedback), is needed to reproduce the early JWST observations at high redshift. However, this does not necessarily imply too early reionization, as previously shown in \citet{2024MNRAS.535L..37M};  instead, it points to an extended reionization history that concludes around $z \sim 6$, as captured by the \eorphisf model.

%-------------------------------------------------%
%%%%%%%%%%%%%%%%%%%%%%%%%%%%%%%%%%%%%%%%%%%%%%%%%%%%%%%%%%%
\section{Conclusion and Discussion} \label{sec:conc}
%%%%%%%%%%%%%%%%%%%%%%%%%%%%%%%%%%%%%%%%%%%%%%%%%%%%%%%%%
We have presented a detailed Bayesian analysis using our semi-analytical framework for modeling the ionizing sources over cosmic time.  Our model incorporates a flexible parameterization of the ionizing photon production rate as a Schechter-like function in halo mass and a power-law in redshift, enabling joint inference on the amplitude, mass scaling, and feedback-driven evolution of ionizing sources. By varying the models' parameters, we have explored a broad range of possible reionization scenarios consistent with a large suite of high-redshift galaxy observations from JWST, along with low-redshift reionization observables such as the ionizing emissivity ($\dot{N}_{\rm ion}$), volume-averaged neutral hydrogen fraction ($x_{\rm HI}$), and the CMB optical depth ($\tau$). This framework provides a coherent way to assess the relative contributions of different galaxy populations and how their evolving emissivities shape the reionization history.

{The weak and strong feedback scenarios identified in our analysis can be understood in terms of the underlying physical processes that regulate star formation and 
ionizing photon production in high-redshift galaxies. In the weak-feedback models (D $= 2.28$), the ionization rate evolves mildly with redshift, consistent with 
the moderate feedback efficiencies derived from radiative transfer simulations \citep{2016MNRAS.457.1550H, 10.1093/mnras/stx420, 2011ApJ...743..169F}. In this regime, stellar feedback (primarily supernova-driven outflows and photoionization heating) scales mildly with redshift, and low-mass halos ($M_h \lesssim 10^{10}\,M_\odot$) dominate the ionizing photon budget, as their star formation is not strongly suppressed. This picture is broadly consistent with the reionization scenario in which numerous faint galaxies collectively drive reionization \citep[e.g.,][]{Finkelstein2019, 2015ApJ...802L..19R}.
In contrast, the strong-feedback models (D $\sim 5$--$7$) require a rapid redshift evolution of $R_{\rm ion}$, which could arise physically from several mechanisms. First, supernova feedback efficiency is expected to increase at 
higher redshifts due to lower metallicities, which reduce radiative cooling losses in supernova remnants and allow the deposited energy to couple more effectively to the surrounding gas \citep{Thornton1998, Kimm2015, Smith2018}. At the same time, higher gas densities and more compact galaxy morphologies at early times can enhance the impact of feedback on the interstellar medium 
\citep{Muratov2015, Hayward2017}. Second, radiative feedback from photoionization heating by the growing UV background suppresses star formation preferentially in low-mass halos ($M_h \lesssim 10^{9}\,M_\odot$) whose shallow potential wells cannot retain photo-heated gas 
\citep{Shapiro1994, Wise2009, 2011ApJ...743..169F, Dawoodbhoy2018, Ocvirk2020}, effectively shifting the dominant sources of ionizing photons toward more massive systems over time.
Third, bursty star formation histories can produce rapid 
UV luminosity variations that mimic stronger effective feedback when time-averaged \citep{Faucher-Giguere2018, Sun2023a, Sun2023b, Pallottini2023}. Fourth, a mass-dependent escape fraction that increases steeply toward higher redshifts for massive halos could also contribute to the effective redshift evolution captured by 
our D parameter. Recent simulations and semi-analytical models find that $f_{\rm esc}$ correlates with halo mass, star formation burstiness, and redshift \citep{Paardekooper2015, 2023MNRAS.521.3077K, Mutch2024, 
2024MNRAS.529.3751C}, and our strong-feedback models' preference for high $f_{\rm esc} \sim 0.5$--$0.6$ is qualitatively consistent with scenarios where feedback-driven channels facilitate photon escape in actively star-forming massive galaxies.
The strong D parameter ($\sim 5$--$7$) thus effectively captures the combined redshift evolution of all these processes which cannot be individually disentangled with our current parameterization. However, the mass dependence, captured by the C parameter, reflects how different halo mass scales contribute to the emissivity budget: higher C 
values favor more massive halos, consistent with scenarios where feedback preferentially suppresses star formation in low-mass systems at early times \citep{Dawoodbhoy2018, Ocvirk2020}. This is also consistent with the `oligarch' reionization scenario proposed by \citet{Naidu2020}, in which a relatively small number of luminous, massive galaxies with high escape fractions dominate the ionizing photon budget, as opposed to the democratic picture favored by the weak-feedback models.}
Our key findings are:

\begin{itemize}
    \item Weak-feedback (\textbf{wf}) {models with dominant contributions} from low-mass galaxies ($M_h \lesssim 10^{10}\,M_\odot$), and weaker redshift evolution/feedback (e.g. $C \sim 0.15, D = 2.28$), reproduce JWST UVLF in $z<10$ but fails to capture the elevated UVLF measurements at $z>10$. In contrast, strong-feedback (\textbf{sf}) models, with higher contribution from massive galaxies ($M_h \gtrsim 10^{10}\,M_\odot$) and stronger redshift evolution/feedback (e.g. $C \gtrsim 0.6$–$1.1, D \sim 5.3$–$6.8$), reproduce the JWST UVLF at higher redshifts, particularly in the range $z \sim 10$-$14$, consistent with the shallow decline observed in the number density of UV-bright galaxies in JWST surveys \citep[e.g.,][]{2023MNRAS.523.1036B, 2023MNRAS.520.4554D, 2024ApJ...969L...2F}, but over-estimate the bright end at $z\leq9$. Additionally, the model constrained by only reionization observables (\textbf{EoR}) {tends to} over-predict the faint-end, showing discrepancies with the JWST UVLF observations, specifically up to $z\sim10$ (see Figure~\ref{fig:uvlf}).
    
    \item  All models calibrated jointly to EoR and JWST observables reproduce the globally averaged quantities, including the evolution of $\rho_{\rm UV}, \rho_{\rm SFR}$, along with $\dot{N}_{\rm ion}, x_{\rm HI}(z)$, and $\tau$, within current observational uncertainties. Interestingly, despite yielding similar reionization endpoints, different model families predict significantly different reionization timelines. Models that match the shape and evolution of the high-redshift JWST UVLF (\eorphisf) favor earlier and more gradual reionization. Notably, even in this model, where reionization begins as early as $z \sim 16$, due to its extended reionization, the integrated optical depth is {\it not} in tension with CMB constraints, resolving the photon budget crisis highlighted by \citet{2024MNRAS.535L..37M}. Other models including the weak feedback models (\textbf{wf}) tend to favor later and more rapid transitions. Nevertheless, all models complete reionization by $z \sim 6$, disfavoring very late scenarios such as those proposed by \citet{2021MNRAS.508.1853B}, and aligning more closely with recent IGM-based constraints \citep[e.g.,][]{2022MNRAS.514...55B, 2023MNRAS.525.4093G} (see Figure~\ref{fig:all_global_cons}).
    
\end{itemize}

These results are based on our current $R_{\rm ion}$ prescription, which depends on halo mass and redshift, though alternative parameterizations incorporating additional physical quantities may also be viable. For instance, some recent JWST measurements of the high-redshift UVLF suggest that a double power-law (DPL) form provides a better fit than the traditional Schechter function \citep[e.g.,][]{2023MNRAS.518.6011D, 2023MNRAS.523.1009B} which indicates a deviation of the exponential cutoff at low mass typically assumed in halo mass - UV luminosity relations. Consequently, exploring a DPL-like parameterization of $R_{\rm ion}$ using the simulations of \citet{2018MNRAS.480.2628F} may improve the agreement with observations. In addition, constructing an emissivity model using simulations that are consistent with current JWST luminosity functions~\citep[e.g. $\rm FIREbox^{HR}$,][]{2025MNRAS.536..988F} may provide new insights into the roles of massive and faint galaxies during reionization. We leave exploring various ways to parameterize $R_{\rm ion}$ and their impact on reionization observables to future works.  

As mentioned earlier, our current model does not explicitly account for dust extinction. However, recent JWST and ALMA observations, along with theoretical studies, indicate that while dust is minimal in most galaxies at $z \gtrsim 10$ \citep[e.g.,][]{2024MNRAS.529.4087T, 2024MNRAS.531..997C}, it can be non-negligible for some systems by $z \sim 7$-$8$, potentially affecting UV luminosities and inferred SFRs \citep{2019MNRAS.487.1844M, 2023Natur.621..267W, 2025MNRAS.544.1502B}. A more detailed treatment of dust attenuation will be incorporated in the future work.

Our findings underscore the importance of the robustness of the joint-inference approach. As JWST continues to expand its reach toward fainter and rarer galaxies, future constraints on the faint-end slope, star formation duty cycles, and feedback-regulated star formation will be crucial to further refine our models of reionization. Integrating these datasets with IGM tomography and upcoming 21-cm observations with SKAO \citep{2009IEEEP..97.1482D, 2019arXiv191212699B}, we expect significantly improved constraints on the feedback-regulated physics of high-redshift galaxies. The wide-field capabilities of the Nancy Grace Roman Space Telescope \citep{2015arXiv150303757S} will further enhance this picture by providing unprecedented statistics on bright and intermediate-luminosity galaxies across large cosmic volumes, enabling precise measurements of galaxy clustering, luminosity functions, and the environments of early ionizing sources.  

\section*{Acknowledgments}
We are grateful to Enrico Garaldi, Massimo Ricotti, Guochao Sun, Aaron Smith and Harley Katz for insightful discussions and suggestions. We thank the anonymous
referee for comments that significantly improved this paper.
AB acknowledges the financial support from the CTC fellowship at the University of Maryland. The computations were run on the Zaratan cluster at the University of Maryland. 

\software{
Astropy \citep{2013A&A...558A..33A}, Colossus \citep{2018ApJS..239...35D}, CoReCon \citep{2023JOSS....8.5407G}, emcee \citep{2013PASP..125..306F}, Matplotlib \citep{2007CSE.....9...90H}, Numpy \citep{2020Natur.585..357H}, SciPy \citep{2020NatMe..17..261V}, GetDist \citep{2025JCAP...08..025L}
}

\section*{DATA AVAILABILITY}
The data underlying this work will be shared upon reasonable request to the corresponding author.

%%%%%%%%%%%%%%%%%%%%%%%%%%%%%%%%%%%%%%%%%%%%%%%%%%%%%%%%%
\bibliography{sample631}{}
\bibliographystyle{aasjournal}
%%%%%%%%%%%%%%%%%%%%%%%%%%%%%%%%%%%%%%%%%%%%%%%%%%%%%%%%%
\appendix
\renewcommand{\thefigure}{\thesection.\arabic{figure}}
\renewcommand{\thetable}{\thesection.\arabic{table}}
\setcounter{figure}{0}
\setcounter{table}{0}
%%%%%%%%%%%%%%%%%%%%%%%%%%%%%%%%%%%%%%%%%%%%%%%%%%%%%%%%%
%
\section{Constraining to UV luminosity function vs. UV luminosity density} \label{appen:uvlf_rho}
\renewcommand{\arraystretch}{1.5} % 1.0 = default, >1 = more spacing
\begin{deluxetable*}{lccccc}
%\tabletypesize{\scriptsize}
\tablecaption{Best-fit parameters.}
\label{tab:phi_rho_val}
\tablehead{
    \colhead{Model} & \colhead{$\rm log_{10}(A/M_{\odot}^{-1}\, s^{-1})$} & \colhead{$\rm log_{10}(B/M_{\odot})$} & \colhead{$\rm C$} & \colhead{$\rm D$} & \colhead{Constraints}
}
\startdata
$\phi$ & $35.114^{+0.798}_{-0.803}$ & $7.398^{+1.672}_{-1.635}$ & $0.487^{+0.023}_{-0.022}$ & $6.160^{+0.162}_{-0.160}$ & $\phi_{\rm UV}$ \\[0.7ex]
$\rho$ & $33.618^{+3.556}_{-2.476}$ & $7.906^{+1.285}_{-1.691}$ & $1.061^{+0.472}_{-0.533}$ & $6.183^{+1.807}_{-2.062}$ & $\rho_{\rm UV}$ \\[0.7ex]
\enddata
\tablecomments{Since $\phi_{\rm UV}$ and $\rho_{\rm UV}$ trace the non-ionizing UV continuum, which is not affected by the ionizing escape fraction, we set $f_{\rm esc}=1$ in this analysis.}
\end{deluxetable*}
\begin{figure}[ht]
    \centering
    \includegraphics[width=0.48\textwidth]{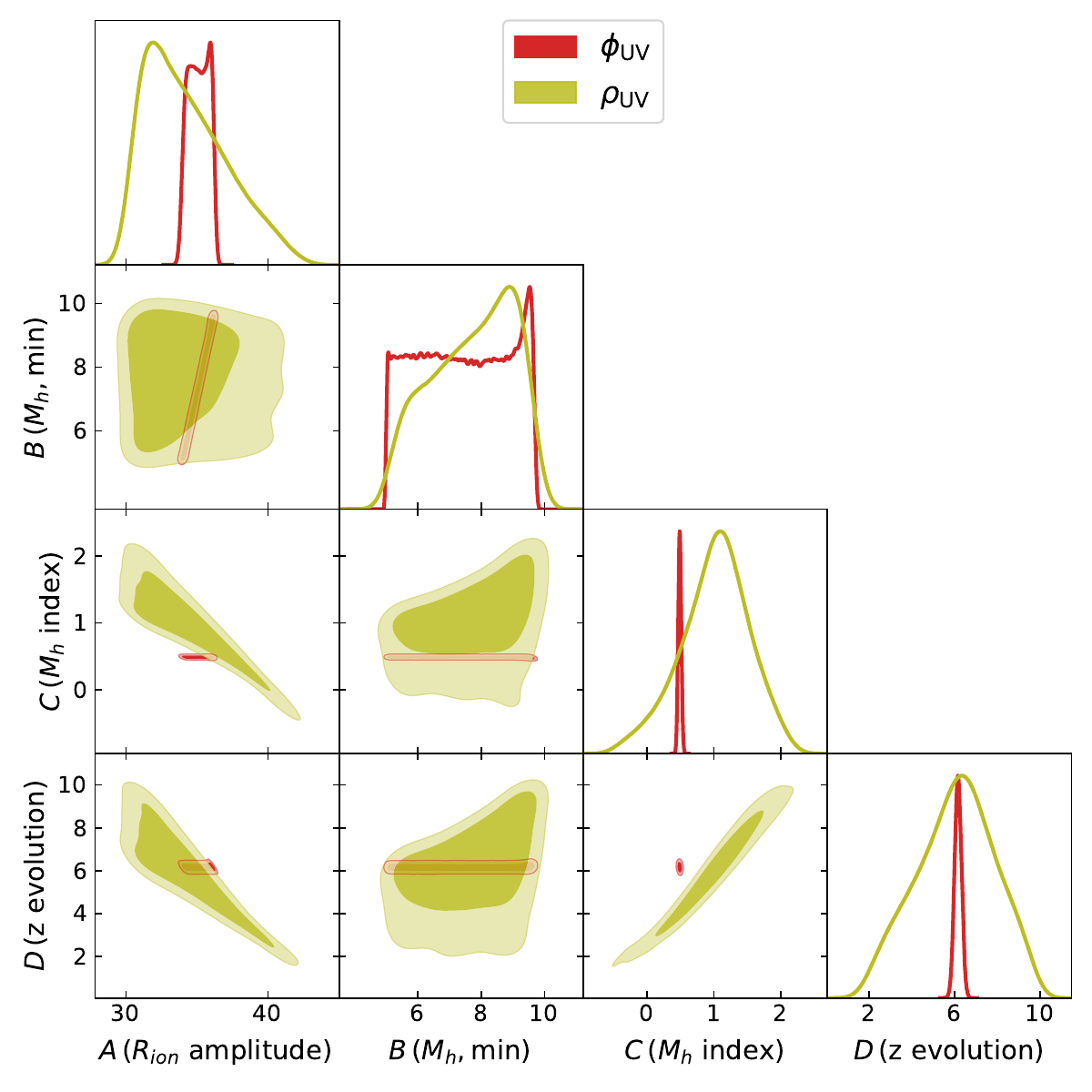}
    \caption{
    Corner plots showing the inferred parameter distributions for using observational constraints, $\phi_{\rm UV}$ and $\rho_{\rm UV}$ separately, allowing all parameters to vary.
    }
    \label{fig:phi_rho_corner_plots}
\end{figure}
Before constraining our models to all observational constraints, we first use the $\phi_{\rm UV}$ and $\rho_{\rm UV}$ observations separately to constrain our source model $R_{\rm ion}$ parameters and examine their correlations. 

The corner plot in the left panel of Figure~\ref{fig:phi_rho_corner_plots} displays the joint and marginal posterior distributions of the four model parameters: A (emissivity amplitude), B ($M_{\rm h, min}$), C ($M_{\rm h}$ index), and D ($z$-evolution, i.e., feedback strength). These parameters are constrained using the JWST UV luminosity function, $\phi_{\rm UV}$, and the UV luminosity density, $\rho_{\rm UV}$ observations, as described in \S\ref{subsubsec:data_uvlf} and \S\ref{subsubsec:data_rhouv}, respectively. This figure illustrates the constraining power of fitting to redshift-dependent non-integrated quantity, $\phi_{\rm UV}$ (in olive) compared to fitting to $\rho_{\rm UV}$ {(in red), where the red} contours are approximately ten times smaller than the olive contours (based on the ratio of the standard deviations of the parameter estimates, see Table~\ref{tab:phi_rho_val}). Both observables, $\phi_{\rm UV}$ and $\rho_{\rm UV}$, weakly constrain the minimum halo mass scale $M_{\rm h}$ (parameter B). Consequently, we fix $\log_{10} B = 7.67$ for subsequent analyses, consistent with the assumption in \citet{2023ApJ...959....2B}, which also approximately corresponds to the midpoint of the values obtained from the $\phi_{\rm UV}$ and $\rho_{\rm UV}$ constraints. The amplitude (A)~–~minimum mass scale (B) plane shows a slight positive correlation, indicating that as the mass cutoff increases, a higher emissivity amplitude is required to reproduce the same observable. A much stronger positive correlation is observed between the feedback strength (D) and the mass index (C), suggesting that an increase in the mass index -- associated with higher emissivity in more massive halos -- requires stronger redshift evolution (feedback) to counterbalance the contribution from low-mass halos. 
Negative correlations are found between the amplitude (A) and both the mass index (C) and feedback strength (D), implying that a reduction in amplitude is necessary to offset the effects of stronger feedback and a steeper $R_{\rm ion}$–$M_{\rm h}$ slope. 

Overall, $\phi_{\rm UV}$ provides tighter individual parameter constraints, whereas $\rho_{\rm UV}$, being an integrated quantity over the entire galaxy population, tends to amplify parameter degeneracies. While the UVLF offers greater constraining power, the resulting best-fit parameters remain within $\pm 1$-$\sigma$ of each other, leading to similar overall effects. However, it is unclear whether this similarity persists once additional observables are incorporated. We therefore include both the \textbf{$\rm EoR$-$\phi$} and \textbf{$\rm EoR$-$\rho$} models in our analysis.

%%%%%%%%%%%%%%%%%%%%%%%%%%%%%%%%%%%%%%%%%%%%%%%%%%%%%%%%%%%
\end{document}